%% file: Romanino.tex
\documentclass{cernyrep}

\usepackage{amsmath}
\usepackage{amssymb}
\usepackage{color}
\usepackage{graphicx}
\usepackage[colorlinks=true, pdftex]{hyperref}

\usepackage{url}

\newcommand{\TeV}{\,\mathrm{TeV}}
\newcommand{\GeV}{\,\mathrm{GeV}}
\newcommand{\MeV}{\,\mathrm{MeV}}
\newcommand{\keV}{\,\mathrm{keV}}
\newcommand{\eV}{\,\mathrm{eV}}
\newcommand{\cm}{\,\mathrm{cm}}

\newcommand{\sm}{{\text{SM}}}

\newcommand{\fracwithdelims}[4]{\left#1 \frac{#3}{#4} \right#2}
\newcommand{\ord}[1]{\mathcal{O}\left( #1 \right)}

\newcommand{\dm}[1]{{\Delta m^2_{#1}}}
\newcommand{\Fig}[1]{Fig.~\ref{fig:#1}}

\newcommand{\eq}[1]{eq.~(\ref{eq:#1})}
\newcommand{\eqs}[1]{eqs.~(\ref{eq:#1})}

\newcommand{\nohyphens}%
        {\hyphenpenalty=10000\exhyphenpenalty=10000\relax}

\DeclareMathOperator{\im}{Im}
\DeclareMathOperator{\re}{Re}

\DeclareMathOperator{\diag}{Diag}

\DeclareMathOperator{\sign}{sign}

\newcommand{\GSM}{G_\text{SM}}

\newcommand{\nuless}{0\nu2\beta}

\newlength{\myem}
\newcommand{\sep}[1]{#1}
\newcounter{mysubequation}[equation]
\renewcommand{\themysubequation}{\alph{mysubequation}}
\newcommand{\mytag}{\stepcounter{mysubequation}%
\tag{\theequation\protect\sep{\themysubequation}}}
\newcommand{\globallabel}[1]{\refstepcounter{equation}\label{#1}}

\renewcommand{\abstractname}{\large\bfseries Abstract}

\newcommand{\SISSA}{SISSA/ISAS and INFN Trieste, Italy}

\begin{document}

\include{Romanino_body}

\end{document}

%% file: Romanino_body.tex
\title{Neutrino Physics}
\author{Andrea Romanino}
\institute{\SISSA}
\maketitle


\begin{abstract}
These lectures aim at providing a pedagogical overview of neutrino physics. We will mostly deal with standard neutrinos, the ones that are part of the Standard Model of particle physics, and with their standard dynamics, which is enough to understand in a coherent picture most of the rich data available. After introducing the basic theoretical framework, we will illustrate the experimental determination of the neutrino parameters and their theoretical implications, in particular for the origin of neutrino masses. 
\end{abstract}



\section{Introduction}

Neutrino physics has played a crucial role in particle physics since the birth of the theory of weak interactions, but the advances in a field requiring the detection of such an elusive particle, have been characterized by long time scales until 1998. After about 70 years of slow (but steady) progress, the findings of the Super-Kamiokande (SK) experiment in 1998 triggered an impressive acceleration and a renewed interest in the field. 

There are various reasons for the interest in neutrino physics. First of all, after decades in which the interpretation of neutrino experiments testing neutrino transitions has been plagued by the uncertain knowledge of the initial fluxes, the SK experiment started an era in which the data interpretation has been relatively clean, with measurements either relatively independent of the flux uncertainties or based on quite a precise knowledge of the fluxes. 

On the theoretical side, the evidence of small but non-vanishing neutrino masses represents one of the few clear indications of physics beyond the standard model of particle physics (SM). The latter predicts in fact vanishing neutrino masses, unless supplemented by additional degrees of freedom or effective interactions (which can be anyway considered as strong hints of new degrees of freedom living at a higher energy scale). 

Moreover, neutrinos play a crucial role in cosmology, another field which is witnessing a fast expansion. They could be responsible for the origin of the baryon asymmetry of the universe, i.e.\ our very existence, they enter the determination of the spectrum of the cosmic microwave background (CMB), the determination of the large scale structures (LSS) in the universe, the delicate chemical equilibriums determining the light element abundances during big bang nucleosynthesis (BBN). Not to mention astrophysics, where they represent a powerful probe of the dynamics of the Sun, of core-collapse supernovae, of high energy sources, etc. 

Last but not least, neutrinos play an important role in many particle physics models and phenomena, and may allow to indirectly access energy scales otherwise largely out of the reach of (natural and laboratory) particle accelerators. They are for example important ingredients in grand unified theories (GUTs), flavour models, lepton flavour violation, lepton violation, neutrinoless double beta decay. 

In this lectures we will discuss many of the issues mentioned above. After some historical remarks, we will discuss neutrino phenomenology, in particular what do we know, and how, about the neutrino parameters. We will then discuss the theoretical impact of the knowledge on neutrino parameters, in particular the implications on the origin of neutrino masses and of the pattern of neutrino masses and mixings. Useful general references are~\cite{GiuntiWeb,Strumia:2006db}. We will use natural units. 

\section{The birth of neutrino physics}

\begin{figure}
\begin{center}
\includegraphics[width=0.35\textwidth]{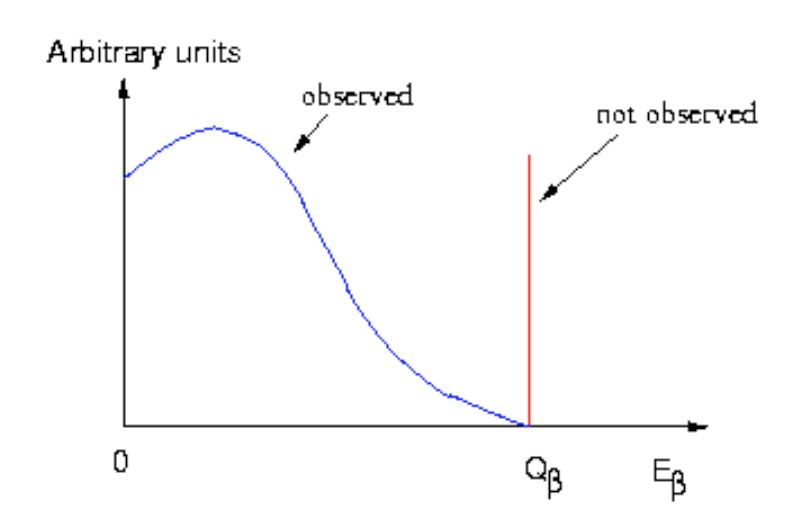}
\raisebox{15pt}[0pt]{\includegraphics[width=0.35\textwidth]{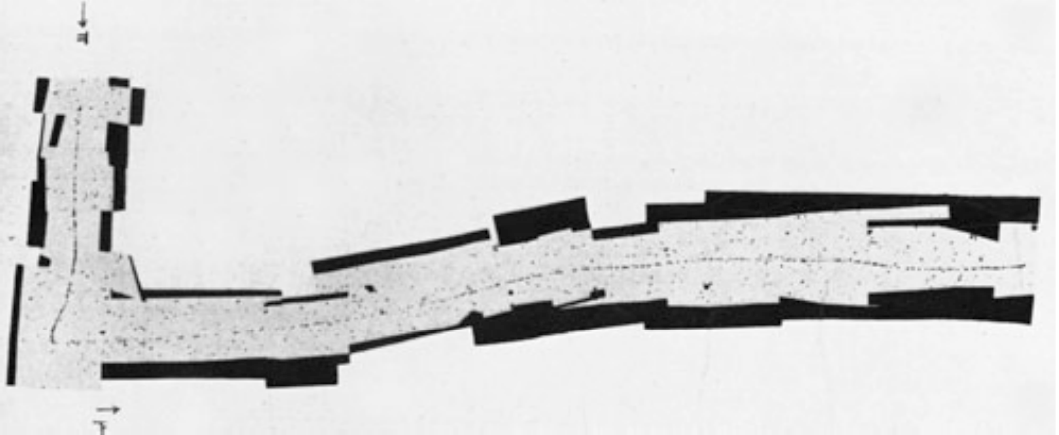}}
\raisebox{-20pt}[0pt]{\includegraphics[width=0.25\textwidth]{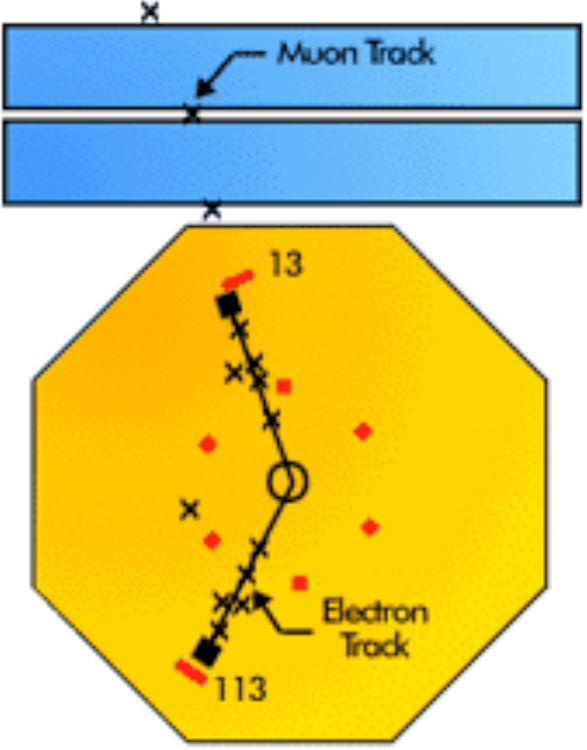}} \\
\end{center}
\hspace{3cm} (a) \hspace{4.5cm} (b) \hspace{4.1cm} (c)
\caption{Neutrino indirect discoveries: (a) electron neutrino from the beta decay spectrum; (b) muon neutrino from the decay of pions from cosmic rays; (c) tau neutrino from $e^+e^- \to e^\pm\mu^\mp X$ anomalous events at SPEAR.}
\label{fig:discoveries}
\end{figure}

The existence of neutrinos has been first postulated by Pauli~\cite{Pauli} in 1930 as a ``desperate remedy'' to save energy conservation in the beta decay of nuclei. In the absence of neutrinos, the two body decay of a nucleus made of $Z$ protons and $A$ nucleons, $(A,Z) \to (A,Z+1) + e^-$, would produce a monochromatic spectrum for the emitted electron. On the other hand, a continuous spectrum, typical of a three body decay, was observed (see \Fig{discoveries}a). Hence Pauli's postulation of the existence of what we now call an electron anti-neutrino, $\overline{\nu}_e$, in the final state of the decay:
\begin{equation}
\label{eq:beta_yrep}
(A,Z) \to (A,Z+1) + e^- + \overline{\nu}_e .
\end{equation}
Fermi took Pauli seriously and in 1934 provided a quantitative description of the phenomenon in terms of an effective interaction that would become the basis of the theory of weak interactions~\cite{Fermi:1934sk,Fermi:1934hr}. His success lead to a wide acceptance of the neutrino hypothesis. The experimental progress in neutrino physics has been quite slow in most of the neutrino history, and it was only in 1956 that Reines and Cowan~\cite{Reines:1953pu,Cowan:1992xc} were able to establish experimentally the existence of the neutrino emitted by nuclear reactors by detecting them through the inverse beta reaction
\begin{equation}
\label{eq:betainverse}
\overline{\nu}_e + p \to n + e^+ .
\end{equation}

The muon neutrino was also introduced to account for missing momentum. Some cosmic ray tracks observed in balloon experiments were in fact exhibiting 90-degrees kinks in correspondence to what would be interpreted as the decay of a charged pion into a muon and a muon neutrino: $\pi^+ \to \mu^+ \nu_\mu$ (see \Fig{discoveries}b).  While such tracks were observed in the late 40's, when the pion was discovered, it was only in 1964 that Lederman, Schwartz, and Steinberger~\cite{Danby:1962nd} detected muon neutrinos by producing the first, prototypical, artificial neutrino beam at Brookhaven. The latter was obtained by sending a beam of protons on target to produce pions, mostly decaying into muons and neutrinos as above. This is still a widely used method to produce neutrinos, in the laboratory and in Nature. The neutrinos were then detected through interactions with matter that were found to produce muons, not electrons, thus indicating that the neutrino detected was not the same as the electron neutrino detected by Reines and Cowan. Moreover, the separate conservation of individual (electron, muon) lepton numbers holds in such processes, once the neutrinos are given the same lepton number as their corresponding lepton. As a consequence, when decaying into an electron, a muon should produce both a muon neutrino and an electron antineutrino: $\mu\to e\, \nu_\mu \overline{\nu}_e$. The fact that muon decay is mostly a three body decay is confirmed by the analysis of the electron spectrum. Another reason why the two neutrinos emitted in the muon decay should be different is that otherwise the muon decay operator would induce a $\mu\to e\gamma$ decay rate much larger than what experimentally allowed. 

Tau neutrinos were associated in 1975 to the tau lepton discovery and detected only recently, in 2000. Tau leptons were discovered at the SPEAR $e^+e^-$ accelerator, which observed anomalous events $e^+e^- \to e^\pm\mu^\mp X$~\cite{Perl:1975bf} (see \Fig{discoveries}c). $X$ represents here one or more invisible particles whose presence in the final state was inferred by the measurement of the missing momentum. The charged particles in the final state are produced by the decay (too fast to be observed) of a $\tau^+\tau^-$ pair produced in the collision. Analogously to the muon case, the $\tau$ lepton should decay into a lighter lepton and a couple of neutrinos, to conserve the individual lepton numbers. The tau neutrinos was thus introduced. Its detection at the DONUT experiment~\cite{Kodama:2000mp} at FNAL was achieved by producing a tau neutrino flux through the $D_s$ decay into $\tau\,\nu_\tau$ and through the challenging observation of the tau produced by the $\nu_\tau$ interaction in matter. 

The obvious question is now whether the story is over or there exists other neutrino species, besides the electron, muon and tau ones. An important constraint on the existence of such additional neutrinos is given by the measurement of the $Z$-boson width at LEP. The decay width depends on the number of (kinematically accessible) decays channels. The measurement agrees with the prediction one obtains taking into account the known charged particles with $m < M_Z/2$ and the three ($2.98\pm0.01$) known neutrinos. The existence of additional neutrinos with $m< M_Z/2$ is therefore excluded. Note that by neutrino here we mean a particle with the same interactions as the three known neutrinos. ``Sterile neutrinos'', hypothetical particles not having any SM gauge interaction, would not contribute to the $Z$ width and are therefore allowed by the $Z$-width constraint. The latter, moreover, translates into a constraint on the number of fermion families. The known fermions are organised in three families with identical gauge quantum numbers, each including a neutrino. The existence of additional families incorporating light ($m< M_Z/2$) neutrinos is therefore also excluded. 

Neutrino oscillation have also a long history. They were postulated in 1957 by Pontecorvo~\cite{Pontecorvo:1957cp,Pontecorvo:1957qd} who, in analogy with $K^0$-$\overline{K}^0$ oscillations, considered neutrino-antineutrino oscillations. The possibility of mixing among electron and muon neutrinos was then considered by Maki, Nakagawa, and Sakata~\cite{Maki:1962mu} in 1962, in analogy to the Gell-Mann hypothesis of quark mixing. The first experimental evidence of neutrino oscillations came in 1968, when Davis, using and experimental technique suggested by Pontecorvo~\cite{Chalk,Pontecorvo:1946mv}, observed a deficit of about 50\% in the measured solar neutrino flux~\cite{Davis:1968cp,Cleveland:1998nv} with respect to what predicted by Bahcall~\cite{Bahcall:1997ha} a few years before. Such an evidence was however plagued by the uncertainties on the theoretical prediction of the solar neutrino flux. As we will see, it was only relatively recently that the SNO experiment was able to get rid of such uncertainties and confirm both the deficit and Bahcall's prediction. 

\section{Neutrino parameters}

Before discussing what we know about them and how, let us define the neutrino parameters. In order to put the discussion in context, we start by describing the theoretical framework and by illustrating the difference between Dirac and Majorana neutrino. 

Most neutrino experiments are characterized by energies much lower than the electroweak scale, $v = 174\GeV$. At such scales, the electroweak symmetry is badly broken and it is convenient to describe the dynamics of the particles light enough to be produced as initial or final states by means of an effective hamiltonian that does not involve heavy fields and obeys the unbroken QED and QCD symmetries (U(1)$_\text{em}$ and SU(3)$_c$ respectively). 

Neutrinos do not couple to photons (QED) nor gluons (QCD). Their interactions are described by an effective lagrangian generated by $W$ and $Z$ exchanges at the electroweak scale. Such a four fermion interaction was first introduced by Fermi and its detailed form was spelled out later ~\cite{Gamow:1936jk,Sudarshan:1958vf,Feynman:1958ty}:
\begin{equation}
\label{eq:4fermion_yrep}
\mathcal{L}_{E\ll M_{Z}}^\text{eff} =
\mathcal{L}_\text{QED+QCD} +4\,\frac{G_F}{\sqrt{2}}\, 
j_c^{\mu} j_{c\mu}^\dagger +
\text{N.C.} + \ldots ,
\end{equation}
where ``N.C.'' denotes the neutral current term, which is not relevant in the following considerations, and the charged current is given by
\begin{equation}
\label{eq:jcc}
j^\mu_c = 
\overline{\nu_{e_{i}}} \,\gamma^\mu P_L \, e_{i} + \overline{u_{i}}\, \gamma^\mu P_L \, d_{i}
\end{equation}
in terms of the three charged lepton Dirac fields $e_i$ ($e_1 \equiv e$, $e_2 \equiv \mu$, $e_3 \equiv \tau$), the three neutrino ``flavour eigenstates'' $\nu_{e_i}$, $i=1,2,3$, the three up quarks $u_i$ ($u_1 \equiv u$, $u_2\equiv c$, $u_3 \equiv t$), the three down quarks $d_i$ ($d_1 \equiv d$, $d_2 \equiv s$, $d_3 \equiv b$). $P_L = (1-\gamma_5)/2$ is the projector on left-chirality fields, $\psi_L \equiv P_L \psi$, and  $P_R = (1+ \gamma_5)/2$ is the projector on right-chirality fields, $\psi_R \equiv P_R \psi$. In the massless limit, left-chirality fields are associated to left-handed helicity particles and right-handed helicity antiparticles. In the massive limit this is not the case. 

Neutrinos are allowed to have non-vanishing masses $m_1$, $m_2$, $m_3$. We denote by $\nu_i$, $i=1,2,3$, the neutrino mass eigenstates fields, which by definition diagonalize the mass matrix. The flavour eigenstates, $\nu_{e_i}$, $i=1,2,3$, diagonalize the charged current. The neutrino flavour eigenstates can be expressed in terms of the mass eigenstates through a $3\times3$ unitary matrix $U$ called Pontecorvo Maki Nakagawa Sakata (PMNS) matrix. The matrix is unitary because it has to preserve the canonical form of the kinetic term of the neutrino fields. As in the case of the CKM matrix describing quark mixing, not all the 9 parameters parameterizing the matrix $U$ are physical. The physical parameters are three mixing angles and, depending on the Dirac or Majorana nature of the neutrinos (see below), one or three CP-violating phases. The standard parameterization of the PMNS matrix in the case of Majorana neutrinos is 
\begin{equation}
\label{eq:PMNS}
U = \left(
\begin{array}{ccc}
c_{12} c_{13} &
s_{12} c_{13} &
s_{13} e^{-i\delta} \\
-s_{12}c_{23}
-c_{12}s_{23}s_{13}e^{i\delta} &
c_{12}c_{23} 
-s_{12}s_{23}s_{13} e^{i\delta} &
s_{23}c_{13} \\
s_{12}s_{23}
-c_{12}c_{23}s_{13}s^{i\delta} &
-c_{12}s_{23}
-s_{12}c_{23}s_{13}s^{i\delta} &
c_{23}c_{13}
\end{array}
\right)
\begin{pmatrix}
1 & 0 & 0 \\
0 & e^{i\alpha} & 0 \\
0 & 0 & e^{i\beta}
\end{pmatrix} ,
\end{equation}
where $s_{ij} = \sin\theta_{ij}$, $c_{ij} = \cos\theta_{ij}$, $\theta_{12},\theta_{23},\theta_{13}$ are three mixing angles, $\delta$ is a CP-violating phases that is physical in both the Majorana and Dirac neutrino cases, and $\alpha,\beta$ are two CP-violating phases that are physical only in the case of Majorana neutrinos and are therefore sometimes called Majorana phases. In the case of Dirac neutrinos, the standard parameterization of the PMNS matrix is given by the first factor only in the RHS \eq{PMNS}. 

From a qualitative point of view, Dirac and Majorana neutrinos differ as follows. In the case of Dirac neutrinos, neutrino and antineutrino are two different particles, each with two possible values of the helicity, for a total of 4 degrees of freedom. Also, the neutrino and antineutrino fields are independent and the neutrino field, as all Dirac fields, splits into two independent components with definite chirality (value of $\gamma_5$): $\nu = \nu_L+\nu_R$. A mass term for Dirac neutrinos does not break lepton number. 

In the case of Majorana neutrinos, the neutrino particle coincides with its antiparticle\footnote{In the case of massless neutrinos, the two helicities do not mix and can be associated to two massless fermions, the neutrino and the antineutrino, with one degree of freedom each. In the case of massive neutrinos, the two helicities are mixed by the mass term (they are both part of the same irreducible representation of the Poincar\'e group) and represent a single particle.}, which gives a total of 2 degrees of freedom. The neutrino and antineutrino fields are not independent (they are related by a matrix transformation) and only have one chirality: $\nu = \nu_L$. A mass term for a single Majorana neutrino field necessarily breaks lepton number and any U(1) charge associated to the neutrino. 

In the massless limit, the distinction between Dirac and Majorana neutrinos is irrelevant. Indeed, the $\nu_R$ component of the Dirac field, if it exists, does not have in this case any interaction, gauge or Yukawa, nor it is mixed by a mass term to the $\nu_L$ component. Therefore, it does not affect the dynamics of the fields produced in the experiments. As a consequence, telling Majorana from Dirac neutrinos in experiments in which the energy is much larger than the neutrino mass (so that we approach the massless limit) is difficult. In particular, the distinction between Majorana and Dirac neutrinos is irrelevant in oscillation experiments and in most other neutrino experiments except when lepton number violation plays a role, as in the case of neutrinoless double beta decay (see Section~\ref{sec:beyond}). 

From a pragmatic point of view, what above is what needed for the comprehension of this and the next Section. In the next subsection we will give additional theoretical details on the nature of neutrino masses that will be mostly needed as a background to Section~\ref{sec:theo} only and can be omitted at a first reading. 

\subsection{The neutrino mass term}
\label{sec:massterm}

In this interlude we would like to discuss in greater theoretical detail the form of the neutrino mass term, the difference between Dirac and Majorana neutrinos, and how the neutrino parameters arise. This requires a basic knowledge of quantum field theory. 

The charged fermions are described by Dirac spinors, four component complex fields. From the point of view of Lorentz transformations, Dirac spinors are not elementary, though. For example, the  electron field, $e$, contains two independent components that have different (inequivalent) Lorentz transformations, characterized by their chirality: $e = e_L + e_R$. In order to be able to write the most general Lorentz invariant mass term or interaction, it is useful to list the fields with equivalent Lorentz transformations. Each Standard Model charged fermion field, and each conjugated field, decomposes into a left and a right component: $u_i = u_{iL}+u_{iR}$, $\overline{u_i} = \overline{u_{iL}} + \overline{u_{iR}}$, $d_i = d_{iL}+d_{iR}$, $\overline{d_i} = \overline{d_{iL}} + \overline{d_{iR}}$, $e_i = e_{iL}+e_{iR}$, $\overline{e_i} = \overline{e_{iL}} + \overline{e_{iR}}$, where $i=1,2,3$ is the family index and I have omitted the color index of quarks. Only the left-handed component of the neutrino field has been observed so far, therefore we do not include a right-handed component in the list for the time being: $\nu_i = \nu_{iL}$, $\overline{\nu_i} = \overline{\nu_{iL}}$. Note that the conjugated of a right-chirality field also has left-chirality. The left-chirality fields are therefore $e_{iL}$, $u_{iL}$, $d_{iL}$, $\overline{e_i}_L = \overline{e_{iR}}$, $\overline{u_i}_L = \overline{u_{iR}}$, $\overline{d_i}_L = \overline{d_{iR}}$, $\nu_{iL}$. The conjugated fields have all right-chirality. The most general Lorentz invariant gauge transformation can in principle mix all the left-chirality fields. Once the gauge transformations have been defined, the most general mass term is given by
\begin{equation}
\label{eq:massterm}
\frac{m_{ij}}{2} \psi_{iL}\psi_{jL} + \text{h.c.} ,
\end{equation}
where $\psi_{iL}$ denotes a generic left-chirality field (the 7 (per family)  fields listed above, in the case of the SM), a proper Lorentz invariant contraction of the Lorentz indexes is understood, and the mass matrix $m_{ij}$ is symmetric and should be invariant under gauge transformation. It is then easy to write the most general mass term for the fermions above. In the effective theory we are considering, the relevant gauge symmetries are the QED and QCD ones, U(1)$_\text{em}$ and SU(3)$_c$. Under SU(3)$_c$ transformations, each left-chirality quark field $u_{iL}$, $d_{iL}$ transforms as a triplet and each left-chirality antiquark field, $\overline{u_{iR}}$, $\overline{d_{iR}}$, transforms as an anti-triplet (leptons are of course invariant). Under U(1)$_\text{em}$ transformations, each left chirality fermion transforms according to its electric charge: $Q_{u_L} = 2/3$, $Q_{d_L} = -1/3$, $Q_{e_L} = -1$, $Q_{\nu_L} = 0$, $Q_{\overline{u_R}} = -2/3$, $Q_{\overline{d_R}} = 1/3$, $Q_{\overline{e_R}} = 1$. Note that the neutrino field does not feel either QCD or QED interactions, hence its elusiveness. It is then an easy exercise to write the most general mass term in the form in \eq{massterm} and invariant under QED and QCD transformations:
\begin{equation}
\label{eq:QEDQCDmassterm}
m^U_{ij} \overline{u_{iR}} u_{jL} + 
m^D_{ij} \overline{d_{iR}} d_{jL} +
m^U_{ij} \overline{e_{iR}} e_{jL} + 
\frac{m^L_{ij}}{2} \nu_{iL}\nu_{jL} + \text{h.c.} . 
\end{equation}
A few comments are in order. The charged fermions mass terms couple two independent fields (that can be combined into a Dirac field). Such mass terms are called ``Dirac'' mass terms. The factor 1/2 is missing there because, for example, $(m^U_{ij}/2) \overline{u_{iR}} u_{jL} + (m^U_{ji}/2) u_{iL} \overline{u_{jR}} = m^U_{ij} \overline{u_{iR}} u_{jL}$, where we have used the fact that the mass term, when written in the form in \eq{massterm}, is symmetric. The neutrino mass term, on the other hand, involves the same set of fields. Such a mass term is called a ``Majorana'' mass term. Any charge carried by the $\nu_{iL}$ (such as total lepton number, for example) is violated by such a mass term. That is why the neutrinos are the only fermions in the above list for which a Majorana mass term is allowed: they are the only neutral fields (under the gauge symmetries we are considering). Note that the observed smallness of neutrino masses, compared to all the other fermion masses, is not explained at this level: the QED and QCD gauge symmetries allow a mass term for both the charged fermions and the neutrinos. We will see in Section~\ref{sec:theo} that a natural explanation for the smallness of neutrino masses arises once the whole SM gauge symmetry $\GSM = \text{SU(3)}_c\times \text{SU(2)}_L\times \text{U(1)}_Y$ is considered. 

If neutrinos have a right-chirality component $\nu_R$ (also uncharged under QED and QCD interactions), the most general mass term for neutrinos becomes
\begin{equation}
\label{eq:nuLnuR}
\frac{m^L_{ij}}{2} \nu_{iL}\nu_{jL}  \to
\frac{m^L_{ij}}{2} \nu_{iL}\nu_{jL} +\frac{m^R_{ij}}{2} \nu_{iR}\nu_{jR} +
m^N_{ij} \overline{\nu_{iR}}\nu_{jL} .
\end{equation}
The first two terms are Majorana, while the third one is a Dirac mass term. The Dirac limit, in which lepton number is conserved if $\nu_L$ and $\nu_R$ are given the same lepton number as $e_L$ and $e_R$, is obtained for $m^L = m^R = 0$. We will see that the right-chirality components of neutrinos, if present, are allowed to get a mass term much heavier than the electroweak scale. This may account for their absence from the effective lagrangian we are considering and for the smallness of the light neutrino masses. For the time being we stick to the economical and theoretically appealing case in which only left-chirality neutrinos are present. 

Fermion masses and mixings are obtained when writing the lagrangian in terms of mass eigenstates. In order to do that, it suffices to determine new linear combinations of the initial fields, 
\globallabel{eq:transformations}
\begin{align}
d'_{iL} &= U^{d_L}_{ij}d_{jL}, & d'_{iR} &= U^{d_R}_{ij}d_{jR}, & 
u'_{iL} &= U^{u_L}_{ij}u_{jL}, & u'_{iR} &= U^{u_R}_{ij}u_{jR}, \mytag \\
e'_{iL} &= U^{e_L}_{ij}e_{jL}, & e'_{iR} &= U^{e_R}_{ij}e_{jR}, & 
\nu'_{iL} &= U^{\nu_L}_{ij}\nu_{jL} , \mytag
\end{align}
such that: i) the kinetic term for the new field is still canonical and ii) the mass terms can be written as
\begin{equation}
\label{eq:masses}
\begin{split}
m^U_{ij} \overline{u_{iR}} u_{jL} + 
m^D_{ij} \overline{d_{iR}} d_{jL} +
m^U_{ij} \overline{e_{iR}} e_{jL} + 
\frac{m^\nu_{ij}}{2} \nu_{iL}\nu_{jL} + \text{h.c.} =  \\
m_{u_i} \overline{u'_{iR}} u'_{iL} + 
m_{d_i} \overline{d'_{iR}} d'_{iL} +
m_{e_i} \overline{e'_{iR}} e'_{iL} + 
\frac{m_{i}}{2} \nu'_{iL}\nu'_{iL} + \text{h.c.} = \\
m_{u_i} \overline{u'_i} u'_i + 
m_{d_i} \overline{d'_i} d'_i + 
m_{e_i} \overline{e'_i} e'_i + 
\left(
\frac{m_{i}}{2} \nu'_{iL}\nu'_{iL} + \text{h.c.}
\right) .
\end{split}
\end{equation}
Because of the first requirement, the mixing matrices $U$ must be unitary. Because of the second one, they should satisfy
\begin{equation}
\label{eq:diag}
m^D = 
U_{d_R}^\dagger m^D_{\text{diag}} 
U^{\phantom{\dagger}}_{d_L}, 
\quad
m^E = 
U_{u_R}^\dagger m^U_{\text{diag}} 
U^{\phantom{\dagger}}_{u_L} ,
\quad
m^E = 
U_{e_R}^\dagger m^E_{\text{diag}} 
U^{\phantom{\dagger}}_{e_L} ,
\quad
m^L = 
U^T_\nu m^\nu_{\text{diag}} 
U^{\phantom{\dagger}}_{\nu},
\end{equation}
where the diagonal matrices have non-negative eigenvalues. It turns out that given generic complex matrices $m^{U,D,E}$ and given a symmetric complex matrix $m^\nu$, it is always possible to find unitary matrices $U^{d_L,d_R,u_L,u_R,e_L,e_R,\nu_L}$ satisfying \eqs{diag}. We can then express the whole lagrangian in terms of the primed fields with definite mass, and drop the primes for convenience. The QED and QCD gauge lagrangian do not change form, as the transformations in \eqs{transformations} conserve the gauge quantum numbers. The only change arises in the charged current, which in terms of the mass eigenstate fields, becomes
\begin{equation}
\label{eq:CC}
j^\mu_c = V_{ij} \overline{u_{iL}} \gamma^\mu d_{jL} + U^\dagger_{ij} \overline{\nu_{iL}}\gamma^\mu e_{jL} ,
\end{equation}
where $V = U_{u_L} U^\dagger_{d_L}$ is the Cabibbo Kobayashi Maskawa (CKM) quark mixing matrix and $U = U_{e_L} U^\dagger_{\nu_L}$ is the PMNS lepton mixing matrix. 

Not all the parameters in $V$, $U$ are physical. Let us consider the CKM matrix first. It is possible to write $V = \diag(e^{i\gamma_1}, e^{i\gamma_2}, e^{i\gamma_3}) V_\text{standard} \diag(1, e^{i\alpha}, e^{i\beta})$, where $V_\text{standard}$ is in the form of the first factor in the RHS of \eq{PMNS}. Moreover, it is possible to reabsorb the phases $\gamma_i$, $i=1,2,3$, $\alpha,\beta$ through a redefinition of the left-handed fields $u_{iL}\to u_{iL}e^{i\gamma_i}$, $d_{2L}\to d_{2L} e^{-i\alpha}$, $d_{3L}\to d_{3L} e^{-i\beta}$. This allows to write $V$ in the standard form $V_\text{standard}$. On the other hand, the field phase transformations used to bring $V$ in the standard form, introduce phases in the mass terms in the second line of \eq{masses}. In order to get rid of them once and for all, it is possible to redefine the phases of the right-chirality fields $d_{iR},u_{iR}$. This shows that the phases $\gamma_i$, $\alpha$, $\beta$ are not physical, as they can be completely eliminated from the lagrangian. In the lepton sector the story would be exactly the same if the neutrinos had a Dirac mass term $m^N_{ij} \overline{\nu_{iR}} \nu_{iL}$ as the charged fermions. On the other hand if, as we assume, the neutrino mass term is Majorana, the phases $\alpha$ and $\beta$ in $U = \diag(e^{i\gamma_1}, e^{i\gamma_2}, e^{i\gamma_3}) U_\text{standard} \diag(1, e^{i\alpha}, e^{i\beta})$ end up being physical. Indeed, while it is possible to eliminate those phases from $U$ by redefining $\nu_{2L}\to \nu_{2L} e^{-i\alpha}$, $\nu_{3L}\to \nu_{3L} e^{-i\beta}$, this transformation would move those phases in the neutrino Majorana mass term. As the latter does not involve an independent field whose phase can be rotated to eliminate $\alpha,\beta$ once and for all, the phases $\alpha,\beta$ turn out not to be unphysical. They are called Majorana phases. 

\subsection{Physical lepton mass and mixing parameters (Majorana neutrinos)}

Let us now go back to phenomenology. As we have seen, the physical mass and mixing parameters in the lepton sector are the 6 charged lepton and neutrino masses and the 6 mixing parameters
\begin{equation}
\label{eq:parameters}
m_e, \;
m_\mu, \;
m_\tau, \quad
m_{1}, \;
m_{2}, \;
m_{3}, \quad
\theta_{23}, \;
\theta_{12}, \;
\theta_{13}, \;
\delta, \quad
\alpha, \;
\beta .
\end{equation}
The physical ranges of the above parameters are $m_{e,\mu,\tau,1,2,3} \geq 0$, $0\leq \theta_{23,12,13} \leq \pi/2$, $0 \leq \delta< 2\pi$, $0 \leq\alpha,\beta < \pi$. 

One important remark concerns the ordering of the neutrino mass eigenstates. In the charged lepton sector (and in the quark sector), the mass eigenstates are ordered with their masses: $m_{e_1} < m_{e_2} < m_{e_3}$. In the case of neutrinos, the convention used is different. By definition, we call $\nu_1$ and $\nu_2$ the two neutrinos whose masses are closer in value, with $m_{1} < m_{2}$. The third mass eigenstate has a larger separation in mass from $\nu_1$ and $\nu_2$, but can be heavier or lighter. If $m_{3} > m_{1,2}$, we say that the neutrinos have a ``normal'' hierarchy. If $m_{3} < m_{{1,2}}$, we say they have an inverse hierarchy. Let us call
\begin{equation}
\label{eq:dm}
\dm{ij} \equiv m^2_j - m^2_i .
\end{equation}
Then we have, by definition, $\dm{12} > 0$ and $0 < \dm{12} < |\dm{23}|$. Corresponding to the two possible hierarchies, $\dm{23}$ can have both signs: $\dm{23} > 0$ in the case of normal hierarchy and $\dm{23} < 0$ in the case of inverse hierarchy. 

Neutrino oscillation phenomena do not depend on the absolute values of neutrino masses but only on the squared mass differences in \eq{dm}. It is then useful to use the following set of lepton mass and mixing parameters, equivalent to the one in \eq{parameters}: 
\begin{equation}
\label{eq:parameters2}
m_e, \;
m_\mu, \;
m_\tau, \quad
\dm{12}, \;
|\dm{23}|, \;
\sign(\dm{23}), \;
\theta_{23}, \;
\theta_{12}, \;
\theta_{13}, \;
\delta, \quad
m_\text{lightest}, \;
\alpha, \;
\beta .
\end{equation}
The neutrino masses $m_{1}$, $m_{2}$, $m_{3}$ have been traded for the equivalent set of parameters $m_\text{lightest}$, the lightest neutrino mass, and $\dm{12}$, $|\dm{23}|$, $\sign(\dm{23})$. The mass parameters $\dm{12}$, $|\dm{23}|$, $\sign(\dm{23})$, the mixing angles $\theta_{23}$, $\theta_{12}$, $\theta_{13}$ and the phase $\delta$ are accessible to neutrino oscillation experiments. The absolute scale of neutrino masses, represented by $m_\text{lightest}$,  and the Majorana phases $\alpha,\beta$ (if physical) are not. The third squared mass difference, $\dm{13}$ is obviously not independent of the first two, $\dm{13} = \dm{12} + \dm{23}$. The experiment shows that $\dm{12} \ll \dm{23}$, so that $\dm{13}\approx\dm{23}$. We have then in first approximation only two squared mass differences, which are sometimes called ``solar'' and ``atmospheric'': $\dm{\text{SUN}} \equiv \dm{12}$, $\dm{\text{ATM}} \equiv \dm{23} \approx \dm{13}$. The name refers, as we will see, to the neutrino source that was first used to measure those parameters. Analogous names are sometimes used for the corresponding mixing angles: $\theta_{\text{SUN}} \equiv \theta_{12}$, $\theta_{\text{ATM}} \equiv \theta_{23}$.

The experimental situation is the following. The charged lepton masses are of course well known. The solar and atmospheric squared mass differences, together with the corresponding mixing angles, are also known. There are bounds on $\theta_{13}$ and $m_\text{lightest}$. No information is available at present on $\sign(\dm{23})$, $\alpha$, $\beta$. Before discussing in detail the experimental determination of the neutrino parameters, we summarize the most relevant information available at present: 
\begin{equation}
\label{eq:parsummary}
\begin{gathered}
\begin{aligned}
\dm{\text{ATM}} & \sim 2.4 \times 10^{-3}\eV^2 & 
\theta_{23} & \sim 45^\circ &
&\text{(ATM, K2K, Minos)}  \\
\dm{\text{SUN}} & \sim 0.76 \times 10^{-4}\eV^2 & 
\theta_{12} & \sim 35^\circ &
&\text{(SUN, KamLAND)}  \\
& \qquad \theta_{13} < 7^\circ \; \text{($2\sigma$)} & & &
&\text{(CHOOZ, Minos + ATM, SUN)}
\end{aligned} \\[2mm]
\begin{aligned}
&|m_{ee}| = |U^2_{ei} m_{\nu_i}| < \ord{1} \times 0.4\eV & \quad
&\text{(Heidelberg-Moscow)} \\
&(m^\dagger m)_{ee} = |U_{ei}|^2 m^2_{\nu_i} < (2.2\eV)^2 & \quad
&\text{(Mainz, Troktsk)} \\
&\sum_i m_{\nu_i} < \ord{1} \eV \; \text{(priors)} & \quad
&\text{(Cosmology)} .
\end{aligned}
\end{gathered}
\end{equation}
The experiments from which the information is obtained are also indicated. ``ATM'' and ``SUN'' denote the atmospheric and solar neutrino experiments respectively. $U_{ei}$ denotes the ``1i'' element of the PMNS matrix, which can be expressed in terms of the parameters in \eq{parameters2} through \eq{PMNS}. The bound from cosmology is subject to uncertainties associated to the priors used in the analysis. 

\section{The determination of the neutrino parameters}

At present, most experimental information on the neutrino mass and mixing parameters comes from experiments measuring neutrino transitions, which are by now known to be due to neutrino oscillations. There are also beta decay experiments aiming at a measurement of the absolute scale of neutrino masses, $m_\text{lightest}$; neutrinoless double beta decay experiments, sensitive to lepton number violation (Majorana vs Dirac neutrinos) and, in the case of Majorana neutrinos, to both $m_\text{lightest}$ and the Majorana phases $\alpha,\beta$; and experiments in astrophysics and cosmology, sensitive to different neutrino properties. 

\subsection{The physics of neutrino oscillation experiments}

Neutrino oscillations arise from the misalignment of the neutrino flavour eigenstate fields, $\nu_{e,\mu\tau}$, coupled to the charged leptons in the charged current interactions, and neutrino mass eigenstate fields, $\nu_{1,2,3}$, eigenstates of the free hamiltonian and therefore associated to the free propagation. Such a misalignment, as we have seen, is quantified by the PMNS matrix: $\nu_{e_i} = U_{ih} \nu_h$, $\overline{\nu}_{e_i} = U^*_{ih} \overline{\nu}_h$. The one-particle states relations involve the conjugate matrix elements: $|\nu_{e_i}\rangle = U^*_{ih} |\nu_{h}\rangle$, $|\overline{\nu}_{e_i}\rangle = U_{ih} |\overline{\nu}_{h}\rangle$. Neutrinos are produced by the charged current interactions of charged leptons, typically electrons or muons. They are therefore in a flavour eigenstate, i.e.\ in a coherent superposition of mass eigenstates. Suppose a neutrino $|\nu_{e_i}\rangle = U^*_{ih} |\nu_{h}\rangle$ is produced by the interaction with the lepton $e_i$ and it freely evolves. Let us compute the probability that the neutrino is found after a time $t$ to be a $|\nu_{e_j}\rangle$ neutrino, for example by means of a charged current interaction with the lepton $e_j$. The free evolution of the initial state gives $e^{-iHt}|\nu_{e_i}\rangle = U^*_{ih} e^{-iE_ht}|\nu_{h}\rangle$, where $E^2_h = (p^2 + m^2_{h})$. The probability that the neutrino is found to be a $|\nu_{e_j}\rangle$ neutrino is therefore
\begin{equation}
\label{eq:osc1}
P(\nu_{e_i} \rightarrow \nu_{e_j})
= \left| \langle\nu_{e_j}| e^{-iHt} |\nu_{e_i}\rangle \right|^2 = 
|U_{jh} e^{-iE_ht} U^\dagger_{hi}|^2 \approx
|U_{jh} e^{-i\frac{m^2_{h}}{2E}t} U^\dagger_{hi}|^2 ,
\end{equation}
where we have approximated $E_h \approx p + m^2_{h}/(2E)$, as $E\gg m_\nu$ in all neutrino oscillation experiments, and the time $t$ can be replaced by the length travelled $L$. We have not specified the helicity of the neutrino, as it is not necessary. This is because in all neutrino oscillation experiments the neutrino energy is way larger than its mass, $E \gg m_\nu$. Since the neutrino interaction only involves the left-chirality component, the neutrino produced will be mostly in an left-handed helicity state, whether it is Majorana or Dirac, and helicity flips, either at production, detection, or during propagation, are largely negligible. Moreover, in the $E \gg m_\nu$ limit, the oscillation probabilities do not depend on Majorana phases. The oscillation probabilities satisfy $P(\nu_{e_i} \to \nu_{e_j}) = P(\overline{\nu}_{e_j} \to \overline{\nu}_{e_i})$, because of CPT invariance. If CP is conserved, $P(\nu_{e_i} \to \nu_{e_j}) = P(\overline{\nu}_{e_i} \to \overline{\nu}_{e_j})$, and equivalently $P(\nu_{e_i} \to \nu_{e_j}) = P(\nu_{e_j} \to \nu_{e_i})$ if $T$ is conserved. The total oscillation probability is of course one, $\sum_j P(\nu_{e_i} \to \nu_{e_j}) = 1$. 

It is instructive to consider the simplest case of two neutrino oscillations. Let us then consider the electron and muon neutrinos only. Up to phases redefinitions, their mixing can be described by a real orthogonal $2\times 2$ matrix, i.e.\ a rotation by an angle $\theta$, which gives a simple expression for the oscillation probability:
\begin{equation}
\label{eq:2nuosc}
\begin{aligned}
\nu_e & =
\nu_1\cos\theta
+ \nu_2\sin\theta \\
\nu_\mu & =
-\nu_1\sin\theta
+ \nu_2\cos\theta
\end{aligned} \quad
\Rightarrow \quad
P(\nu_e\rightarrow\nu_\mu)
= \sin^22\theta
\sin^2
\frac{\Delta m^2L}{4E} ,
\end{equation}
where $\dm{} = m^2_2 - m^2_1$ can be taken positive by definition. In order to obtain predictions for the outcome of realistic experiments, the oscillation probability has to be convoluted with the source energy spectrum, the distribution in the position of the neutrino emission and detection, the scattering cross sections, the experimental resolution and efficiency. 

Let us comment on the form of the two neutrino oscillation formula. The oscillation amplitude $A = \sin^22\theta$ is determined by the mixing angle $\theta$ and does not allow to distinguish (in vacuum) the physically inequivalent $\theta$ and $\pi/2 -\theta$ values. The squared mass difference determines the oscillation length $\lambda = 4\pi E/\dm{} \approx 2.48\,\text{km} (E(\text{GeV})/\dm{}(\text{eV}^2))$, or equivalently the oscillation phase $\phi = \dm{} L/(4E) \approx 1.27 (\dm{}(\text{eV}^2) L(\text{km})/E(\text{GeV}))$. In order to determine both the oscillation parameters, it is best to consider an experiment in which the neutrinos travel a distance comparable to their oscillation length, $L\sim\lambda$. In the $L\ll \lambda$ limit, in fact, $P(\nu_e\rightarrow\nu_\mu) \approx \sin^22\theta\, (\Delta m^2L/(4E))^2$ and even a detailed measurement of the oscillation probability as a function of $E$ and $L$ would determine the product $\sin^22\theta\cdot \Delta m^2$ only. In this limit, oscillations have not enough time to occur and the expression for the probability can be obtained in perturbation theory (which represents a check of the correctness of the formula). The oscillation probability is proportional to $(L/E)^2$ and the neutrino flux decreases with the geometrical factor $1/L^2$, therefore the number of neutrino oscillation events measured in a detector is approximately independent of the distance $L$, within this limit. In the $L \gg \lambda$ limit, on the other hand, the oscillations are so fast that they average out and $P(\nu_e\rightarrow\nu_\mu) \approx \sin^22\theta/2 = \sin^2\theta \cos^2\theta + \cos^2\theta \sin^2\theta$. Only the mixing angle can be measured in this limit. The oscillation probability is independent of $E$, $L$ and the number of oscillation events decreases with $1/L^2$. In this ``classical'' limit, the oscillation probability is the sum (over $i$) of the probabilities that the initial neutrino $\nu_e$ is a $\nu_i$ times the probability that the neutrino $\nu_i$ is observed to be a  $\nu_\mu$. 

\begin{figure}
\begin{center}
\includegraphics[width=0.35\textwidth]{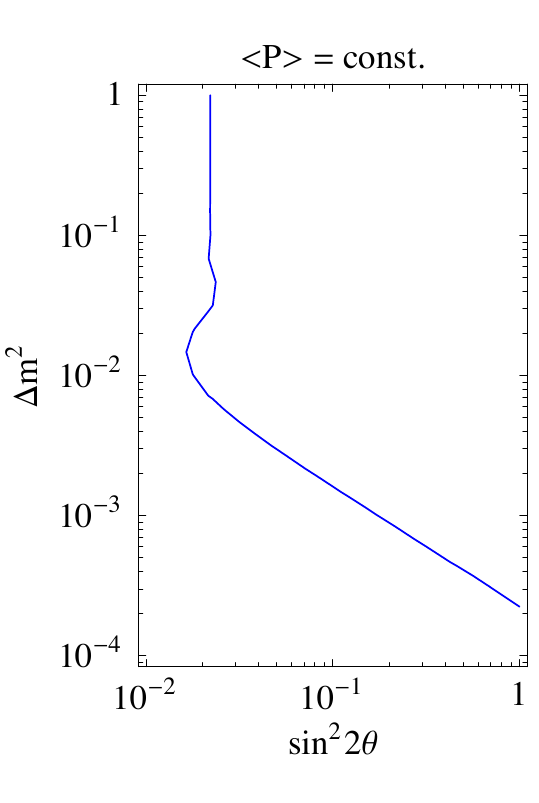}
\end{center}
\caption{Typical sensitivity plot of a neutrino experiment.}
\label{fig:2nuosc}
\end{figure}

The situation is illustrated in \Fig{2nuosc}, where the typical sensitivity plot of a neutrino experiment is plotted. Assuming that the experiment is sensitive to a given (averaged) oscillation probability, the sensitivity in the $\sin^2 2\theta$--$\dm{}$ plane is shown. The two limits considered above can be recognized in the lower and upper part of the plot respectively. In order to measure both $\sin^22\theta$ and $\dm{}$, a measurement of the averaged probability is not enough. The $E$ or $L$ dependence has also to be measured, better if in the $L\sim\lambda$, or $(\dm{}L/(4E)) \sim 1$, regime. 

The above derivation of the oscillation formula is simplistic for a number of reasons. First of all, it holds in vacuum only. The coherent (or incoherent) effect of matter in neutrino propagation can be very important, as we will show below. Moreover, the neutrino coherence assumed in the derivation of the oscillation formula can be lost for a number of reasons, besides the necessary averages mentioned above. Because the wave packets associated to the different mass eigenstates making up a given flavour eigenstate travel at slightly different velocities, for example (this is relevant when the distance travelled is very large). Or because the neutrino production process typically involves at least another particle in the final state. The quantum mechanics reduction to the neutrino subsystem also induces a loss of coherence. Finally, the derivation assumed that the neutrino mass eigenstates are all in a pure eigenstate with same definite momentum. It is sometime argued that it is more appropriate to assume that they have the same energy instead. A proper derivation would take into account the precise form of the density matrix describing the initial state and its momentum distribution, as obtained from the dynamics of the production process. In this context, using the fixed momentum or fixed energy description just amounts to a change of variable in the integration over the momentum distribution of the initial neutrino state. 

Let us consider now the three neutrino case. The exact three neutrino formulas are
\begin{gather}
\begin{aligned}
P(\nu_{e_i}
\rightarrow
\nu_{e_j})
& = P(\overline{\nu}_{e_j}
\rightarrow
\overline{\nu}_{e_i})
= P_{\text{CPC}}
+ P_{\text{CPV}} \\
P(\overline{\nu}_{e_i}
\rightarrow
\overline{\nu}_{e_j})
& = P(\nu_{e_j}
\rightarrow
\nu_{e_i})
= P_{\text{CPC}}
- P_{\text{CPV}}
\end{aligned}
\notag
\\[1mm]
\label{eq:3nuosc}
\begin{aligned}
P_{\text{CPC}} & = 
\delta_{ij}
-4 \re(J^{ji}_{12}) S^2_{12}
-4 \re(J^{ji}_{23}) S^2_{23}
-4 \re(J^{ji}_{31}) S^2_{31} \\
P_{\text{CPV}} & = 
8\sigma_{ij} J_{\text{CP}}
S_{12}S_{23}S_{31}
\end{aligned}
\\[1mm]
S_{hk} 
= \sin
\frac{\dm{hk}L}{4E}, \quad
J^{ji}_{hk} =
U^{\phantom{\dagger}}_{jh}
U^\dagger_{hi}
U^{\phantom{\dagger}}_{ik}
U^\dagger_{kj},
\quad
\im(J^{ji}_{hk})
= \sigma_{ji}\sigma_{hk}
J_{\text{CP}},
\quad
\sigma_{ij} 
= \sum_k \epsilon_{ijk}
= \pm 1, 0 ,
\notag
\end{gather}
where $P_{\text{CPC}}$ and $P_{\text{CPV}}$ are the CP conserving and CP violating parts of the oscillation probability respectively. Note again the independence of the formulas above of Majorana phases and of the absolute neutrino mass scale. 

Despite the existence of three neutrinos, the results of neutrino oscillation experiments are often shown in a two neutrino oscillation context and mainly determine a single mixing angle and squared mass difference. This is because the experimental values of the neutrino parameters are such that often, in first approximation, the three neutrino oscillation formula reduces to a two neutrino one. For example, the CHOOZ experiment, as we will see, measures the probability of electron neutrino disappearance, $P(\nu_e\to\nu_e)$, for $L/E$ values such that the $S_{12}$ terms in \eq{3nuosc} are negligible (because of the small $\dm{12}$). For the same reason $S_{23}\approx S_{12}$, so that we can approximate $P(\nu_e\rightarrow\nu_e) \approx 1
-\sin^22\theta_{13} \sin^2 (\dm{23}L/(4E))$, a two neutrino oscillation formula with $\theta = \theta_{13}$ and $\dm{} = \dm{23}$. This allows the CHOOZ experiment to set an upper bound on $\theta_{13}$~\cite{Apollonio:1999ae} that makes the $\theta_{13}$ contribution negligible, in first approximation, in the solar and atmospheric neutrino experiments. In particular, in atmospheric neutrino experiments, $S^2_{12}\ll 1$, $S^2_{23} \approx S^2_{13}$, $\theta_{13}\ll 1$, so that $P(\nu_\mu\rightarrow\nu_\tau) \approx \sin^22\theta_{23} \sin^2 (\dm{23}L/(4E))$, $P(\nu_e\rightarrow\nu_{\mu,\tau}) \ll 1$, and the results can be interpreted in terms of $\nu_\mu\leftrightarrow\nu_\tau$ oscillations with $\theta = \theta_{23}$ and $\dm{} = \dm{23}$. In solar neutrino experiments, detecting the disappearance of the electron neutrinos produced by the Sun, the $S^2_{23}$ and $S^2_{13}$ terms are suppressed by $\theta_{13}$ and $P(\nu_e\rightarrow\nu_e) \approx 1-\sin^22\theta_{12} \sin^2(\dm{12}L/(4E))$, leading to a determination of $\theta = \theta_{12}$ and $\dm{} = \dm{12}$. 

In the next subsections, we will discuss the experimental determination of neutrino parameters by oscillation experiments. 

\subsection{Experimental determination of $\dm{23}$ and $\theta_{23}$}

\begin{figure}
\begin{center}
\includegraphics[width=0.35\textwidth]{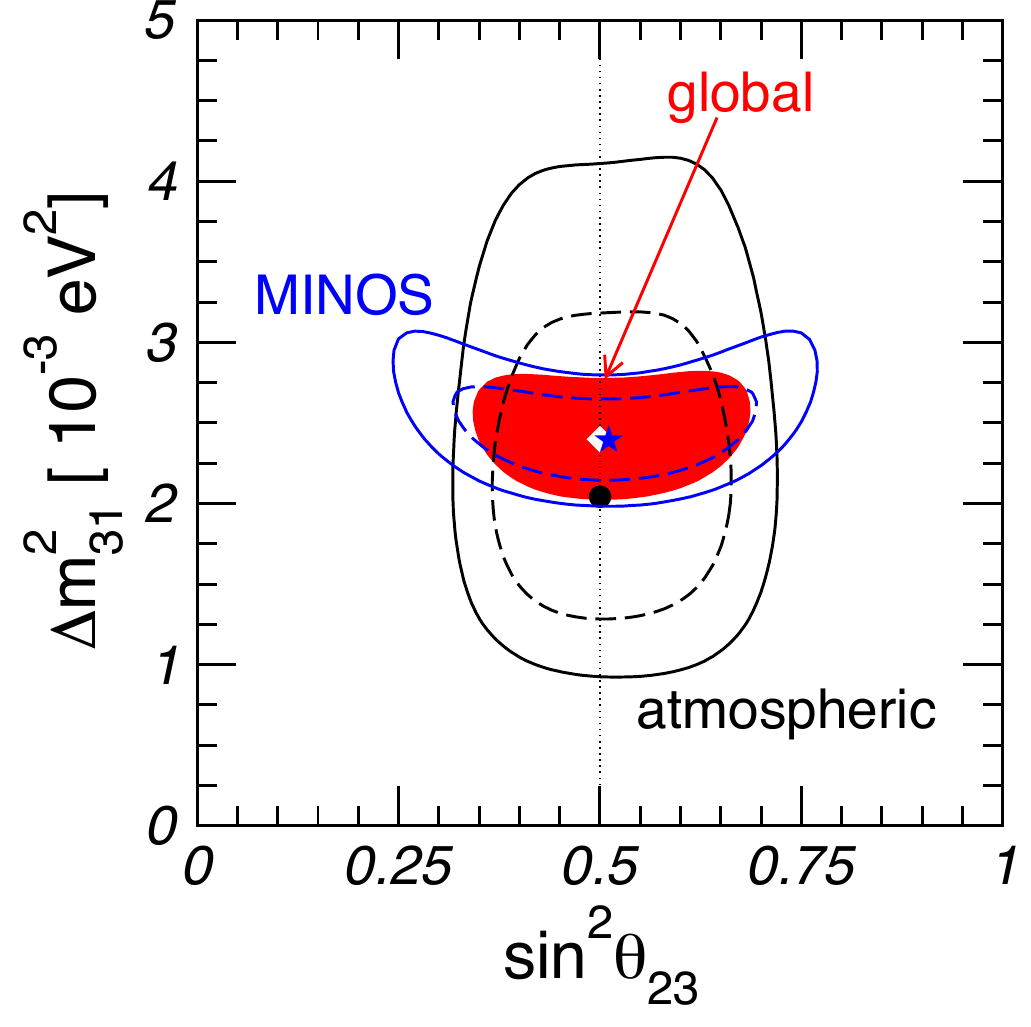}\hspace{1cm}
\includegraphics[width=0.35\textwidth]{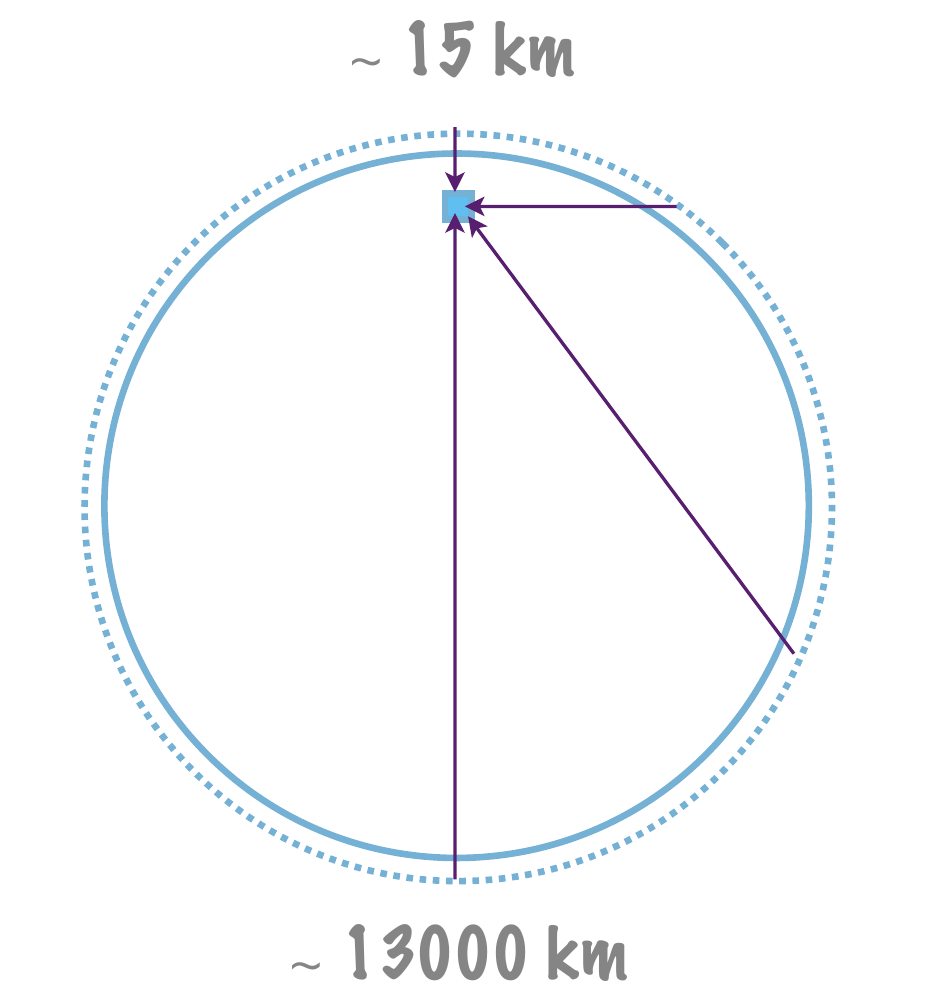}
\end{center}
\hspace{4.7cm} (a) \hspace{5.4cm} (b) 
\caption{Global fit of the $\dm{23}$ and $\theta_{23}$ parameters (a). Schematic representation of the distance travelled by atmospheric neutrinos (b).}
\label{fig:atm}
\end{figure}

The experimental determination of $\dm{23}$ and $\theta_{23}$ is mostly due to the Super-Kamiokande (SK) measurement of atmospheric neutrinos, and to the K2K, Minos, and Opera experiments. The result of a global fit is shown in \Fig{atm}a~\cite{Schwetz:2008er}. Let us discuss the main ingredients entering the above determination.

\subsubsection{Atmospheric neutrinos}

Atmospheric neutrinos arise from cosmic ray interactions in the atmosphere. Charged pion produced by such interactions decay mostly through the decay chain $\pi^+\to\mu^+\nu_\mu \to e^+\nu_e \overline{\nu}_\mu \nu_\mu$ (analogously for negative pions, SK does not tell neutrinos from antineutrinos), thus producing in first approximation two muon neutrinos for each electron neutrino. The ratio of muon to electron neutrinos reaching the Earth is actually slightly larger than two because i) energetic muons have a long life-time and may not decay before reaching the Earth and ii) Kaons are also produced by cosmic ray interactions. Atmospheric neutrinos are detected by experiments placed underground (to shield cosmic rays, but not neutrinos). The neutrinos travel a distance ranging from $10\,\text{km}$ to more than $10^4\,\text{km}$, as shown in \Fig{atm}b. Their energy ranges from 0.1 to 10 and more GeV. The oscillation phase for $\dm{23}$ oscillations is therefore typically $\dm{23}L/(4E) = 10^{-2}$--$10^{2}$, centred around 1, the value we argued is experimentally the best to reveal oscillation. As the neutrino flux produced in the atmosphere is obtained by theoretical simulation characterized by significant uncertainties, the measurement of the absolute muon or electron neutrino flux does not allow to firmly establish the occurrence of neutrino flavour transitions. On the other hand, the measurement of the muon to electron neutrino ratio has a smaller theoretical uncertainty and is therefore more reliable. Even more reliable is the variation of the muon and electron fluxes (and their ratio) with the distance travelled, i.e.\ with the direction (zenith angle) from which they reach the detector. The latter measurements by SK provided in 1998 the first firm evidence of neutrino flavour transitions and opened the modern era of neutrino physics. 

Super-Kamiokande is a large water Cherenkov detector located in the Kamioka mine, in Japan, $2.7\,\text{km}$ underground. It contains about 50 ktons of water and is surrounded by about 13000 photomultipliers. In order to perform the analysis above, a measurement of the neutrino flavour, direction, the energy is needed. Let us see how such information is, at least partial, obtained. 

\begin{figure}
\begin{center}
\includegraphics[width=0.35\textwidth]{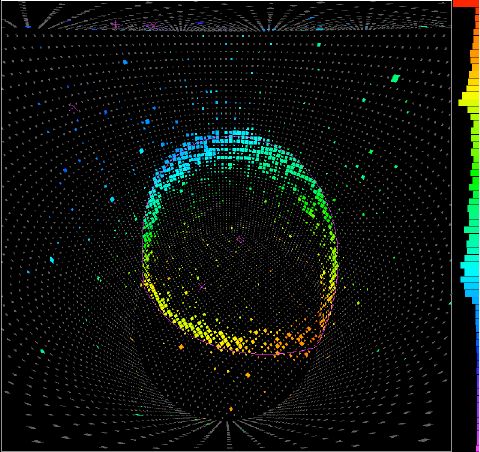}\hspace{1cm}
\includegraphics[width=0.35\textwidth]{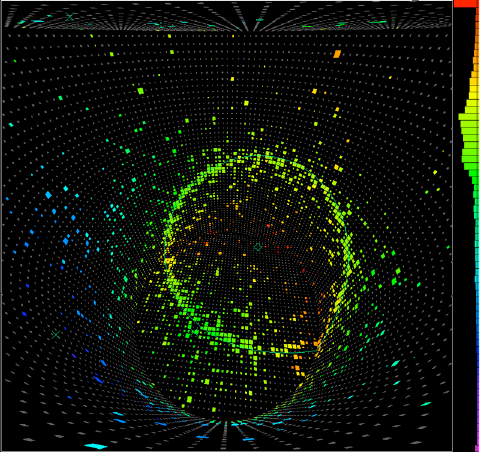}
\end{center}
\hspace{4.2cm} (a) \hspace{5.9cm} (b) 
\caption{Cherenkov rings produced by a muon (a) and an electron (b) in Super-Kamiokande.}
\label{fig:rings}
\end{figure}

Neutrinos can be detected through their charged current interactions with the nuclei: $\nu_{e_i} + N \to e_i + N'$. The lepton $e_i = e,\mu$ produced in the interaction is ultra-relativistic and produces a cone of Cherenkov light while it travels through the water, which is detected by the photomultipliers. When a lepton stops inside the detector the photomultipliers detect a ring of Cherenkov light. The nature of the lepton can be told by the shape of the ring: a muon produces a relatively clean ring, while the ring produced by electrons is more fuzzy, as shown in \Fig{rings}. The position of the ring allows to determine the lepton direction, which is correlated to the neutrino direction if the neutrino energy is larger than about a GeV. If the lepton is produced in the detector and stops inside the detector, its energy can be measured by the amount of Cherenkov light collected by the photomultipliers. The lepton energy is not strongly correlated to the neutrino energy, but it cannot exceed it, which is enough to provide an handle on the energy dependence of the neutrino flavour transition probability. The best events are therefore the ``fully contained multi-GeV'' events. Neutrinos also interact through neutral current interactions with nuclei such as $\nu + N \to \nu + N + \pi^0 \to \nu+N+\gamma\gamma$, which also produce a signal in the detector. Tau leptons can also be produced if the neutrino is energetic enough to exceed the kinematical threshold for production. Taus quickly decay into hadrons, producing a signal similar to the neutral current one. 

By now the statistics accumulated by Super-Kamiokande is impressive. Not only it allows to establish neutrino transitions without any doubt, despite the oscillation pattern is too smeared out by the poor neutrino-lepton energy correlation to be observed explicitly, but it also allows to attribute the transitions to $\nu_\mu \leftrightarrow \nu_\tau$ oscillations. In particular, no depletion of the electron neutrino flux with the distance travelled has been observed, which is compatible with the CHOOZ bound~\cite{Apollonio:1999ae} on $\nu_e$ transitions. Also, oscillations into sterile neutrinos, hypothetical additional neutrinos not feeling any SM gauge interaction, are ruled out or bound to have a marginal role. The same holds for exotic disappearance mechanisms such as neutrino decay, or Lorentz or CPT violation. 

\subsubsection{Accelerator experiments}

The Super-Kamiokande results have been confirmed by a number of experiments using neutrinos produced at accelerators. Opera is a sophisticated detector at the Gran Sasso laboratory in Italy designed to explicitly detect $\nu_\tau$ appearance from a $\nu_\mu$ neutrino beam produced at CERN. Such an appearance would confirm the indirect, but solid, interpretation of the SK results in terms of $\nu_\mu\leftrightarrow\nu_\tau$ oscillations. The tau produced by the $\nu_\tau$ charged current interaction in the detector is observed in emulsion films. Unfortunately, the expected statistics is not very high, but a first candidate $\nu_\tau$ event has been recently reported~\cite{Agafonova:2010dc}. The K2K experiment in Japan used the SK detector to measure the disappearance of $\nu_\mu$ from a pulsed beam produced at KEK. The initial flux is measured by a detector placed near the neutrino source. The average neutrino energy is slightly above $1\GeV$, and the distance travelled by neutrinos is about $250\,\text{km}$, which gives an oscillation phases of order one, as desired. The muon scattering angle in the detector can be measured, together with its energy. The kinematics of the charged current interaction then allows to reconstruct the neutrino energy. The experimental results have been reported in~\cite{Ahn:2006zza}. Another important experiment is Minos, in the Sudan mine, $735\,\text{km}$ north of Fermilab, where the (pulsed) $\nu_\mu$ beam is produced. The average neutrino energy is higher than in K2K, to give again an oscillation phase around one. As in the case of K2K, the initial flux is measured by a near detector. Neutrino interactions in steel are measured in this case by means of a magnetized tracking calorimeter. Minos can observe $\nu_\mu$ charged current events (penetrating muons) and therefore $\nu_\mu$ disappearance, which gives a determination of $\theta_{23}$ and $\dm{23}$ in agreement with the SK one~\cite{Adamson:2011ig}. It can also see neutral current interactions of any neutrino (they produce a diffuse hadron shower), which confirms that oscillations into sterile neutrinos cannot account for the $\nu_\mu$ disappearance~\cite{Adamson:2011ku}. It can detect $\nu_e$ charged current interactions (compact electromagnetic showers) and therefore set a bound on $\nu_\mu\to\nu_e$ oscillations, which translates into a bound on $\theta_{13}$~\cite{Adamson:2010uj} compatible (although at present weaker) with the CHOOZ one. The presence of a magnetic field allows Minos to tell $\mu^+$ from $\mu^-$, which in turn allows to test CP-violation (although with a poor sensitivity). The possibility to switch from a $\nu_\mu$ to a $\overline{\nu}_\mu$ beam allows to test CPT violation. A mild, not very significant, tension between the $\nu_\mu$ and $\overline{\nu}_\mu$ determinations of the oscillation parameters has been recently reported~\cite{Adamson:2011fa}. 

\subsection{Experimental determination of $\dm{12}$ and $\theta_{12}$}

\begin{figure}
\begin{center}
\includegraphics[width=0.35\textwidth]{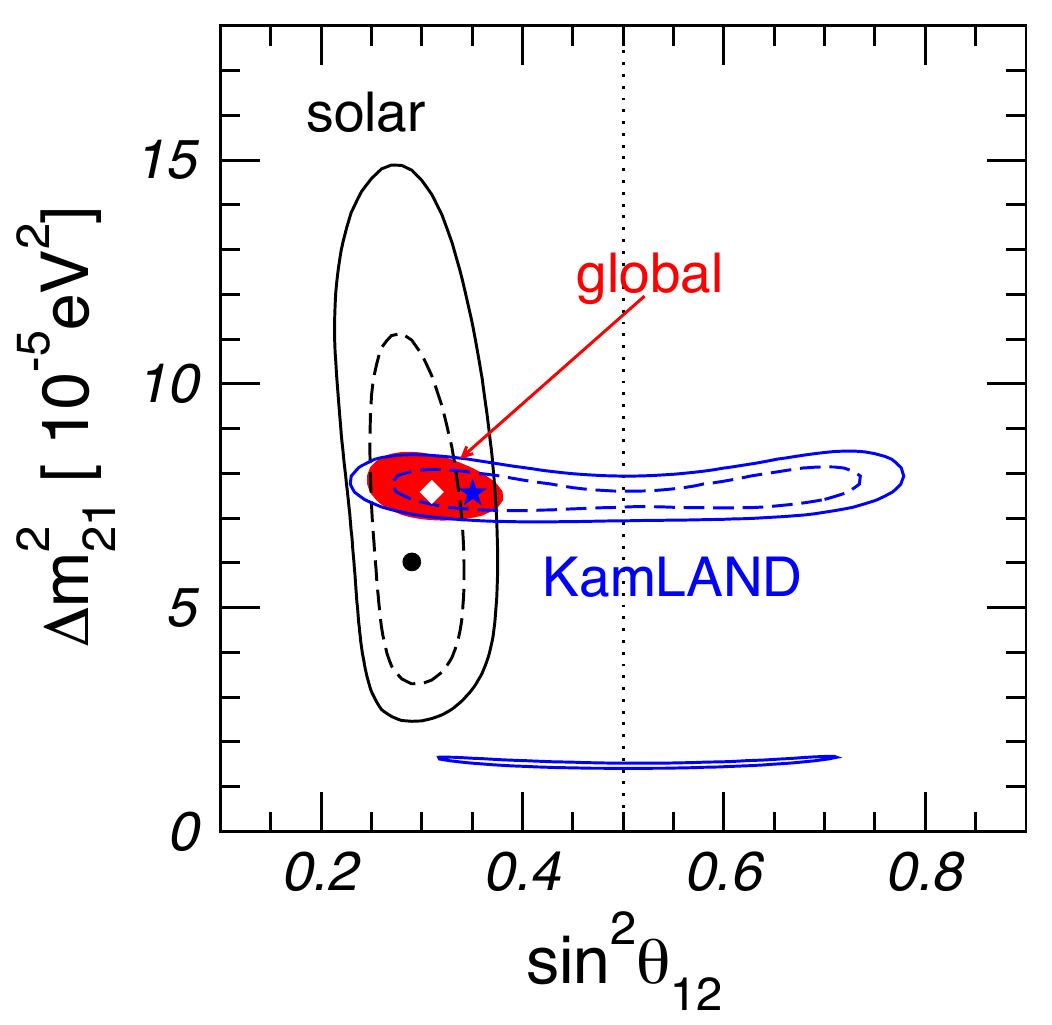}\hspace{1cm}
\raisebox{20pt}[0pt]{\includegraphics[width=0.45\textwidth]{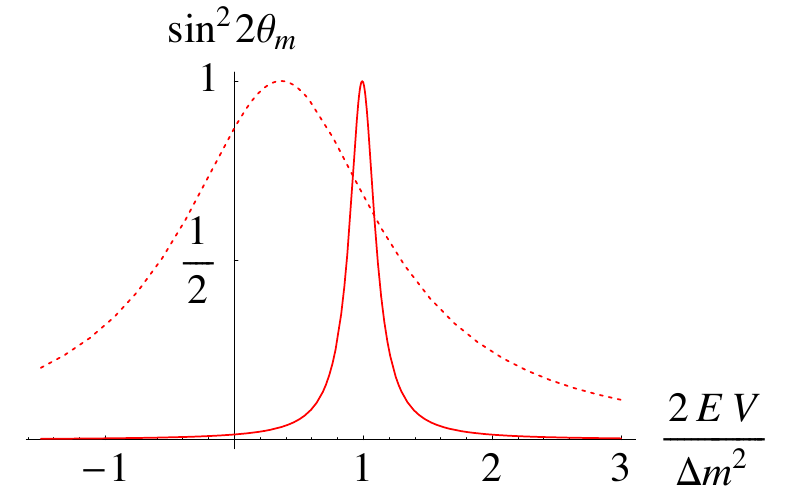}}
\end{center}
\hspace{4cm} (a) \hspace{5.8cm} (b) 
\caption{Global fit of the $\dm{12}$ and $\theta_{12}$ parameters (a). Dependence of the mixing angle in matter with the neutrino energy in the case $\dm{} > 0$ and for two values of the mixing angle in vacuum, $\theta = 0.6$ (dashed curve) and $\theta = 0.06$ (solid curve) (b).}
\label{fig:sun}
\end{figure}

The experimental determination of $\dm{12}$ and $\theta_{12}$ is mainly due to the SK, SNO, Borexino, KamLAND experiments. The result of a global fit is shown in \Fig{sun}a~\cite{Schwetz:2008er}. Let us discuss the main ingredients entering the above determination. In order to illustrate the physics of solar neutrinos, it is necessary to discuss neutrino propagation in matter

\subsubsection{Matter effects in neutrino propagation}

As neutrinos do not feel electromagnetic or strong interactions, they can travel through ordinary matter without experiencing a single scattering interaction. The mean free path of a neutrino in a medium as dense as the Earth's mantle is in fact $\lambda(E) \sim 10^{9}\,\text{km}\, (\text{GeV}/E)^2$ and even in the core of the Sun is $\lambda(E) \sim 10^{10}\,\text{km}\, (10\MeV/E)^2$, both much larger than the distance travelled in the medium. The energy normalization is appropriate for atmospheric and solar neutrinos respectively. Only in extraordinarily dense matter, such as a proto-neutron star core, neutrinos have a mean free path, $\lambda(E) \sim 10\,\text{cm}\, (100\MeV/E)^2$, trapping them in a random walk lasting about 10 seconds. 

This does not mean, however, that matter does not affect neutrino propagation in the Earth and in the Sun. While incoherent scattering is proportional to the square of the weak interaction Fermi coupling suppressing the process, forward coherent scattering~\cite{Wolfenstein:1977ue}, affecting the phase of the neutrino wave function, is proportional to only one power of the Fermi coupling. Let us then compare the rate of incoherent scattering, $dP_\text{sc}/dx$, where $P_\text{sc}$ is the incoherent scattering probability, and the rate of change of the neutrino phase due to coherent forward scattering, $d\phi_\text{co}/dx$:
\begin{equation}
\label{eq:coinco}
\begin{aligned}
&\text{incoherent:} &
& dP_{\text{sc}}/dx
\sim G^2_{\text{F}} E^2 n \\
&\text{coherent:} &
& d\phi_{\text{co}}/dx
\sim G_{\text{F}} n
\end{aligned}
\rightarrow
\frac{dP_{\text{sc}}}
{d\phi_{\text{co}}}
\sim G_{\text{F}}E^2
\sim 10^{-5}
\fracwithdelims{(}{)}{E}{\text{GeV}}^2 ,
\end{equation}
where $n$ is the matter number density and $E$ is the neutrino energy. We therefore see that the coherent effect is largely dominant. While the effect on the neutrino phase would be unobservable in the absence of neutrino oscillations, the impact on oscillations may be significant, as the phases of the three neutrino flavour eigenstates are affected in different ways. 

Coherent scattering in the propagation can be accounted for by adding to the free hamiltonian for the three neutrinos an effective ``MSW''~\cite{Wolfenstein:1977ue,Mikheev:1986gs,Mikheev:1986wj} potential. In the flavour eigenstate basis, the Hamiltonian then reads
\begin{equation}
\label{eq:MSWhamiltonian}
H = 
\frac{1}{2E} U
\begin{pmatrix}
m^2_1 & & \\
& m^2_2 & \\
& & m^2_3
\end{pmatrix}
U^\dagger
+\begin{pmatrix}
V & & \\
& 0 & \\
& & 0
\end{pmatrix}
+\text{universal terms,}
\end{equation}
where $U$ is the PMNS matrix and $V$ is the MSW potential. The three flavour neutrinos feel different potentials in matter, because they have different weak interactions. At the tree level, in neutral matter with no muon or tau lepton number (or with $L_\mu=L_\tau$) and a negligible neutrino density, such as the Earth or Sun matter, $V_\mu = V_\tau$ and $V = V_e - V_\mu = \sqrt{2} G_\text{F} n_e$, where $n_e$ is the electron neutrino number density. The difference between the electron neutrino potential and the muon and tau one is due to the fact that the electron neutrino can interact through charged current interactions with electrons, while the muon and tau neutrinos cannot. The Hamiltonian in \eq{MSWhamiltonian} determines neutrino propagation. In the antineutrino case, $U \to U^*$ and $V\to -V$. Let us see what are the consequences of the presence of the MSW potential are as far as neutrino propagation \emph{in constant density} is concerned. 

The case of constant density is relevant for the neutrino propagation in the Earth, for example. The Earth has a density profile that is not constant but can be in first approximated to be constant both in the mantle, where the mass density is approximately $\rho_m\sim 3$--$5\,\text{g}/\text{cm}^3$, and in the core, where $\rho_c\sim 10$--$15\,\text{g}/\text{cm}^3$. The effect of the Earth in neutrino propagation is important for i) atmospheric neutrinos (only through the subdominant $\nu_e \leftrightarrow\nu_{\mu,\tau}$ transitions, as $\nu_\mu\leftrightarrow\nu_\tau$ transitions are not affected), ii) solar and supernova neutrinos, iii) terrestrial experiments (in the case of a long baseline). 

One of the most interesting consequences of the presence of the MSW term is the possibility of a resonant enhancement of the oscillation amplitude. In order to illustrate such an effect, let us consider a simple two neutrino case, in which the Hamiltonian can be written as
\begin{equation}
\label{eq:MSW2nu}
\renewcommand{\arraystretch}{1.5}
H =
\begin{pmatrix}
\displaystyle
\sin^2\theta
+ \frac{2EV}{\Delta m^2} &
\sin\theta\cos\theta \\
\sin\theta\cos\theta &
\cos^2\theta
\end{pmatrix}
\frac{\Delta m^2}{2E}
+\text{universal terms} .
\end{equation}
It is then clear then a resonant enhancement of the mixing angle takes place when the two diagonal elements coincide. In such a case, the mixing angle ``in matter'' (i.e.\ obtained from the diagonalization of the matrix in \eq{MSW2nu}), $\theta_m$, becomes maximal, $\theta_m = 45^\circ$, no matter how small is the mixing angle in vacuum, and the squared mass difference in matter gets correspondingly suppressed: 
\begin{equation}
\label{eq:enhancement}
\frac{2EV}{\Delta m^2}
= \cos2\theta
\Rightarrow
\left\{
\begin{aligned}
& (\sin2\theta)_m = 1, \; \\
& (\Delta m^2)_m = 
\Delta m^2
\sin2\theta
\end{aligned}
\right. .
\end{equation}
Such a mixing enhancement takes place for an appropriate value of the neutrino energy if $\theta < 45^\circ$ and $V\cdot \dm{} > 0$ or if $\theta > 45^\circ$ and $V\cdot \dm{} < 0$. The dependence of the mixing angle in matter with the neutrino energy is shown in \Fig{sun}b in the case $\dm{} > 0$ and for two values of the mixing angle in vacuum, $\theta = 0.6$ (dashed curve) and $\theta = 0.06$ (solid curve). We see that there exists an energy for which the enhancement is maximal even for small mixing angles, but the energy in which the angle is sizeable is correspondingly small. It is also interesting to follow the dependence of the two Hamiltonian neutrino eigenstates, $(\nu_{1,2})_m$, with the neutrino energy, in the case $\theta\ll 1$, for example (still assuming $V\cdot \dm{} > 0$, so that the resonance does take place). When the neutrino energy is small, $(\nu_{1,2})_m$ coincide with the mass eigenstates $\nu_{1,2}$. Since for $\theta \ll 1$ the electron neutrino is close to $\nu_1$, we have $\nu_e \approx (\nu_1)_m$. In the opposite limit in which $(2EV/\dm{})\gg 1$, the first diagonal element in \eq{MSW2nu} becomes the heaviest and dominates, which means that $\nu_e \approx (\nu_2)_m$. By crossing the resonance, the electron neutrino moves from the first to the second Hamiltonian (propagation) eigenstate. The same effect takes place if the neutrino energy is constant but the matter density, and therefore $V$, varies. Which may play an important role for the neutrino propagation in matter with varying density. 

The precise relation between the mixing angle and squared mass difference in vacuum and in matter is given, in the two neutrino case, by the following formulas:
\begin{equation*}
\label{eq:resformulas}
\sin^22\theta_m
= \frac{\sin^22\theta}
{1+\displaystyle
\fracwithdelims{(}{)}
{2EV}{\Delta m^2}^2
-2\cos2\theta
\frac{2EV}{\Delta m^2}} ,
\quad
(\Delta m^2)_m
= \Delta m^2\left[
1+\fracwithdelims{(}{)}
{2EV}{\Delta m^2}^2
-2\cos2\theta
\frac{2EV}{\Delta m^2}
\right]^{1/2} .
\end{equation*}
We can define a resonant energy $E_\text{res}$ by 
\begin{equation}
\label{eq:Eres}
\frac{2EV}{\Delta m^2}
= \frac{E}{E_{\text{res}}}
\cos2\theta, 
\quad
E_{\text{res}}
= \frac{\Delta m^2}{2V}
\cos2\theta
\approx 8\GeV\left(
\frac{\Delta m^2}
{2\cdot 10^{-3}\eV^2}
\frac{n_e}
{1.65\,\text{gr/cm}^3}
\right) .
\end{equation}
Note also that 
\begin{equation}
\label{eq:boh}
\frac{(\sin^22\theta)_m}
{\sin^22\theta} = 
\fracwithdelims{[}{]}
{\Delta m^2}
{(\Delta m^2)_m}^2 .
\end{equation}
The equations above show that matter effects are negligible when $E\ll E_\text{res}$ or when $L\ll \lambda_m$, where $\lambda_m$ is the oscillation length in matter. In the latter case, in fact, one can approximate $\sin \phi\approx \phi$, where $\phi$ is the oscillation phase. Matter effects then cancel because of \eq{boh}. 

\bigskip

Let us now consider propagation in matter with varying density. Let us still stick to the two neutrino case. The Hamiltonian is time dependent, as the MSW potential varies during the propagation: $H(t) = H_\text{free} + V(t)$. The exact solution for the evolution of the neutrino wave functions are non-trivial and have usually to be obtained numerically. There is however one important case in which the evolution is easy to follow: the adiabatic limit. In such a limit, the variation of the Hamiltonian is much slower than the variation of the oscillation phase. As a consequence, a neutrino which at a certain time is in a given eigenstate of the full Hamiltonian $H(t)$ will remain, in first approximation, in that eigenstate (which however will vary together with $H(t)$). Such an adiabatic evolution takes place if 
\begin{equation}
\label{eq:adiabaticlimit}
\frac{d\theta_m}{dx}
\ll 
\frac{(\Delta m^2)_m}{2E} 
\end{equation}
during the evolution, where $dx$ represents the variation in the neutrino position. 

\begin{figure}
\begin{center}
\includegraphics[width=0.44\textwidth]{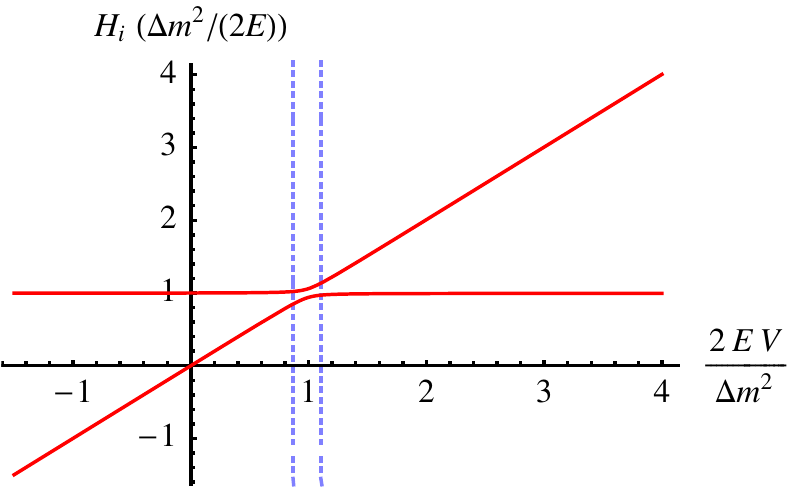}
\raisebox{-20pt}[0pt]{\includegraphics[width=0.55\textwidth]{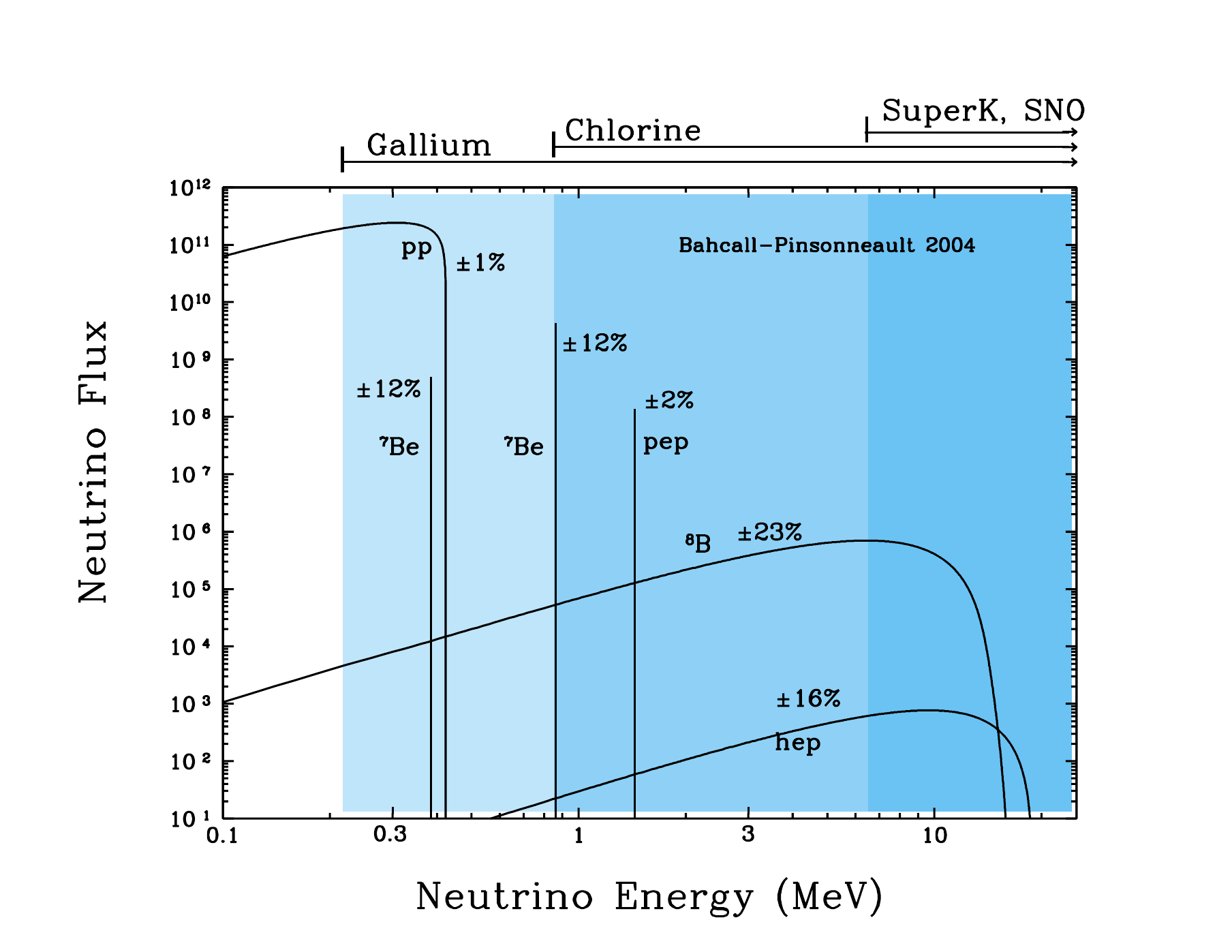}}
\end{center}
\hspace{2.9cm} (a) \hspace{7.4cm} (b) 
\caption{Dependence of the Hamiltonian eigenstates on the MSW potential for $\theta = 0.06$ (a). Contributions to the solar neutrino flux (b). }
\label{fig:MSW}
\end{figure}

An important consequence of the adiabatic evolution is the possibility of large flavour swaps even for small mixing angles. This may happen if the neutrino, while traveling, crosses a resonance because of the variation in the matter density and therefore of $V$. The situation is illustrated in \Fig{MSW}a. There, the dependence of the two Hamiltonian eigenstates (in units of $\dm{}/(2E)$) with $V$ ($2EV/\dm{}$) is shown for a small value of the mixing angle, $\theta = 0.06$. In the relevant case, solar neutrino oscillations, the mixing angle will not be small, but this example better shows how striking the effect can be. Consider the case in which an electron neutrino is emitted in a medium with a density high enough that $2EV/\dm{} \gg 1$ (the inner part of the Sun for example). As we have seen, in such conditions, $\nu_e \approx (\nu_2)_m$. Suppose now the density decreases during the evolution until is vanishes and the evolution is adiabatic. Once the neutrino is out of the medium, it will still be in the second eigenstate of the Hamiltonian. Which in vacuum is $\nu_2 = \nu_e \sin\theta + \nu_\mu \cos\theta$. The probability that the electron neutrino has become a muon neutrino is therefore $P(\nu_e\to\nu_\mu) \approx \cos^2\theta$. The transition probability turns out to be close to one even if the mixing angle is small. Such an effect cannot hold for arbitrarily small mixing angles of course (for $\theta = 0$ there cannot be any effect), which means that the adiabatic approximation must fail when $\theta$ is small enough. This can be seen from \eq{adiabaticlimit}. The adiabatic condition is worse at the resonance, where \eq{adiabaticlimit} becomes
\begin{equation}
\label{eq:adpar}
\gamma\equiv
\frac{\Delta m^2}
{2E(V'/V)_{\text{res}}}
\frac{\sin^22\theta}
{\cos2\theta}\gg1,
\end{equation}
where $V'$ is the derivative of the MSW potential with respect to the position. If $\theta$ is small enough, the adiabatic condition at the resonance is not fulfilled. If the adiabatic condition holds at production and detection and it fails only in a small region around the resonance, the ``level crossing'' probability is given in first approximation by the Landau-Zener formula 
\begin{equation}
\label{eq:LZ}
P(\nu_1\rightarrow\nu_2)
\equiv P_c
\approx e^{-\gamma/2} ,
\end{equation}
where $\gamma$ is the adiabaticity parameter in \eq{adpar}. The Landau-Zener approximation fails in the extreme non-adiabatic regime, $\gamma \ll 1$. 

\subsubsection{Solar neutrinos}

The discussion of the neutrino evolution in varying density applies to solar neutrinos. Solar neutrinos are electron neutrinos produced in the burning process that produces the solar energy: $4p+2e\to {}^4\text{He} + 2\nu_e + 26.7\MeV$. The process takes place through different reactions. Correspondingly, we have different types of solar neutrinos, characterized by different energy spectra. Among them, we have the pp neutrinos, from the $pp\to d e^+\nu_e$ reaction, that have by far the largest flux (which is then well known because it can be derived from a measurement of the total solar luminosity) but have quite a small energy, $E < 0.42\MeV$; the Be neutrinos, from ${}^7\text{Be}+ e \to {}^7\text{Li}+ \nu_e$, with a significant, monochromatic flux with $E = 0.863\MeV$; and B neutrinos, with a small flux, but a more energetic spectrum, extending up to more than $10\MeV$. The latter are the only ones that can be seen by the SK and SNO experiments. The different contributions to the solar neutrino flux are shown in \Fig{MSW}b. 

Several experiments have been devised to measure the solar neutrino flux, starting from the historical Chlorine experiment in the Homestake mine in the US, by Davis~\cite{Cleveland:1998nv}, which gave the first evidence of a neutrino deficit, although with respect to an uncertain theoretical prediction. The latter was a radiochemical experiment. The neutrino reaction $\nu_e\,\mbox{}^{37}\text{Cl} \rightarrow e\,\mbox{}^{37}\text{Ar}$, with energy threshold $E > 0.814\GeV$, was detected by separating the few tens of atoms of Argon produced by chemical methods and by counting them through their beta decay back to the initial isotope. Analogous methods were used in the Gallium experiments (SAGE~\cite{Abdurashitov:1999zd}, at the Baksan lake, Russia and Gallex/GNO~\cite{Hampel:1998xg}, at the Gran Sasso laboratories), exploiting the $\nu_e {}^{71}\text{Ga} \to e\, {}^{71}\text{Ge}$ reaction, with threshold $E > 0.233\MeV$. Such experiments were not able to measure the time at which the reaction happened nor the direction of the incoming neutrinos, but they have the lowest energy thresholds, as shown in \Fig{MSW}b. In particular, the Gallium experiments are the only ones sensitive to pp neutrinos. 

The Super-Kamiokande experiment detects solar neutrinos~\cite{Hosaka:2005um} through elastic scattering with electrons in the water, $\nu_{e,x} \to \nu_{e,x} e$, where $x$ stands for $\mu$ or $\tau$, with an energy threshold $E > 5.5\MeV$. The electron and muon/tau neutrino cross sections are different, with the latter smaller by a factor 6--7, because the charged current interactions do not contribute. 

The Sudbury Neutrino Observatory (SNO) experiment, near Sudbury, Canada, uses heavy water, $D_2O$, and can detect neutrinos through three types of processes. Elastic scattering (ES), $\nu_{e,x} e\to \nu_{e,x} e$, involves all types of neutrinos. Electron and muon/tau neutrinos have different cross sections, however, as in the SK case. If $\Phi_e$ and $\Phi_{\mu+\tau}$ are the electron and muon/tau neutrino flux reaching the Earth, the ES measurement determines $\Phi_e + 0.155 \Phi_{\mu+\tau}$. The neutrino direction can be determined from the electron direction, which allows to tell the solar neutrinos from the background by their direction. Charged current interactions (CC), $\nu_e D \to ppe$, only involve electron neutrinos and therefore determine $\Phi(\nu_e)$. The neutrino energy spectrum can be reconstructed from the electron one. Neutral current interactions (NC), $\nu_x D \to \nu_x pn$, involve all types of neutrinos with equal cross section. They therefore allow to determine the total neutrino flux $\Phi_e + \Phi_{\mu+\tau}$. The SNO experiment has played for solar neutrino a role similar to SK for atmospheric neutrinos, to the extent to which it allowed to obtain a clear evidence of solar electron neutrino transitions, independent of the theoretical uncertainties on the initial neutrino flux. This is because the three reactions, ES, CC, and NC, measure three independent linear combinations of the electron and muon/tau fluxes. It is then possible to determine (and over-constrain) both fluxes, as shown in \Fig{snokam}a~\cite{Aharmim:2006kv}. In particular, the total neutrino flux reaching the Earth (directly given by the NC measurement), barring  exotic phenomena, determines the initial electron neutrino flux. From the experimental point of view, the use of heavy water is necessary in order to obtain neutrino CC interactions (in water only antineutrinos can interact with the proton in the Hydrogen and the neutrino interaction with the neutrons in the Oxygen has a too high thresholds). Chlorine was added in a second phase of the experiment to enhance the neutron capture cross section, which, through the $\gamma$ produced, is an important handle to detect the crucial NC processes~\cite{Aharmim:2009gd}. Adding ${}^3\text{He}$ proportional chambers in the third phase of the experiment further improves the NC detection, as it allows to see the single neutrons. 

The Borexino experiment, at the Gran Sasso laboratories, also uses the elastic scattering process, as SK and SNO, but is sensitive to lower energy neutrinos, in particular to the ${}^7\text{Be}$ ones, as it uses a scintillator detector. Such a measurement~\cite{Arpesella:2008mt} is important as it constrains the electron neutrino survival probability for values of the neutrino energy in which matter effects in the sun are negligible. Moreover, such an experiment was able to measure ``geo-neutrinos''~\cite{Bellini:2010hy}, $\overline{\nu}_e$ from natural radioactivity with $E < 3\MeV$. 

\subsubsection{KamLAND}

\begin{figure}
\begin{center}
\includegraphics[width=0.44\textwidth]{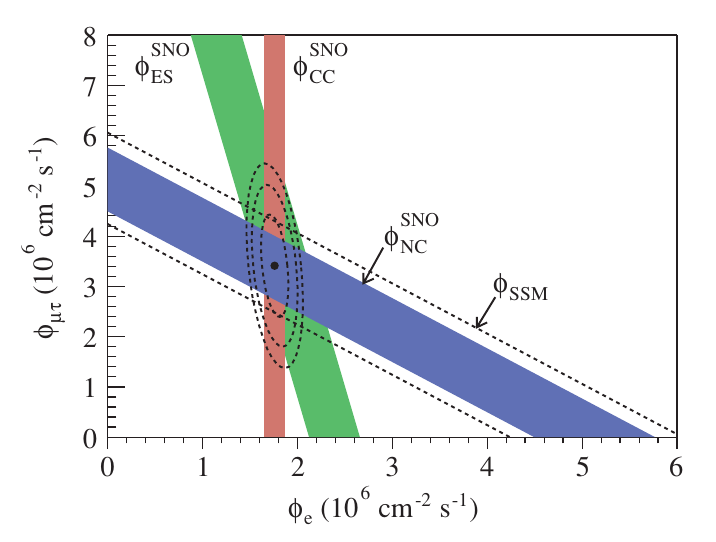}\hspace{0.5cm}
\raisebox{-10pt}[0pt]{\includegraphics[width=0.45\textwidth]{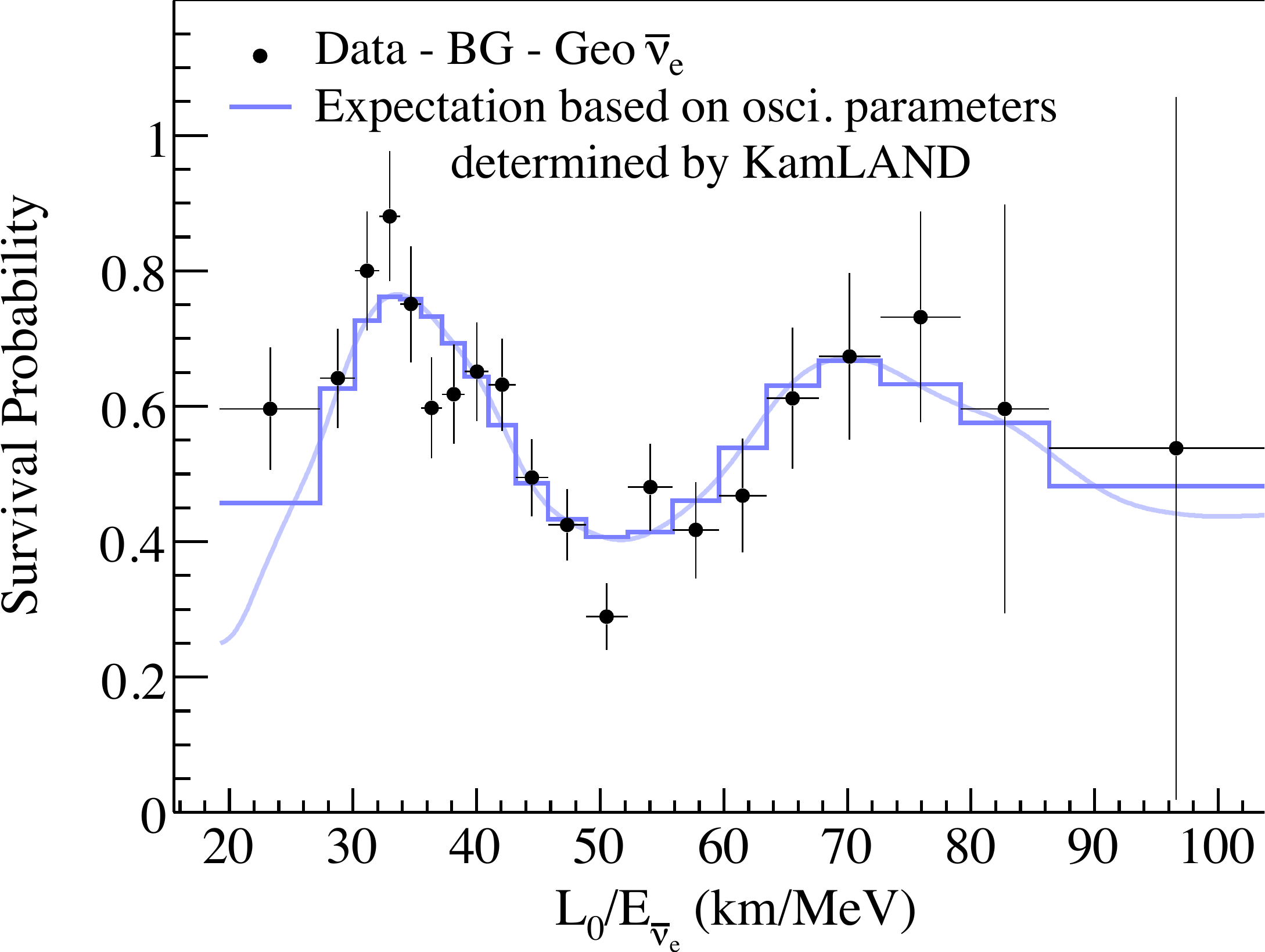}}
\end{center}
\hspace{4.cm} (a) \hspace{7.cm} (b) 
\caption{Determination of the electron and muon/tau neutrino fluxes by SNO (a). Electron antineutrino survival probability at KamLAND (b).}
\label{fig:snokam}
\end{figure}

The Kamioka Liquid-scintillator Anti-Neutrino Detector (KamLAND) experiment, near the Super-Kamiokande experiment, also plays a crucial role in the determination of the $\dm{12}$ and $\theta_{12}$ parameters, as the determination is not obtained by using solar neutrinos, but terrestrial neutrinos ($\overline{\nu}_e$) emitted by several nuclear reactors in Japan. The neutrino energy is of the order of a few MeV, the average distance travelled is about 200$\,$km, giving an order one oscillation phase for the $\dm{12}$ oscillation frequency, $\dm{12}L/(4E) \sim 1$. The detection is performed by means of the CC interaction $\overline{\nu}_e p \to e^+ n$ in the scintillator, with both the electron and the delayed coincidence with the $\gamma$ signal from the neutron capture used to observe it. The neutrino energy is directly related to the positron energy, $E_{\nu_e} = E_{e^+} + m_n - m_p$, which allows to measure the neutrino oscillation probability as a function of the energy and as a consequence i) to obtain a good $\dm{12}$ determination~\cite{:2008ee} and ii) to observe the oscillation pattern, including an oscillation dip, in the survival probability, as shown in \Fig{snokam}b~\cite{:2008ee}. 

\subsection{The unknown oscillation parameters}

\begin{figure}
\begin{center}
\includegraphics[width=0.44\textwidth]{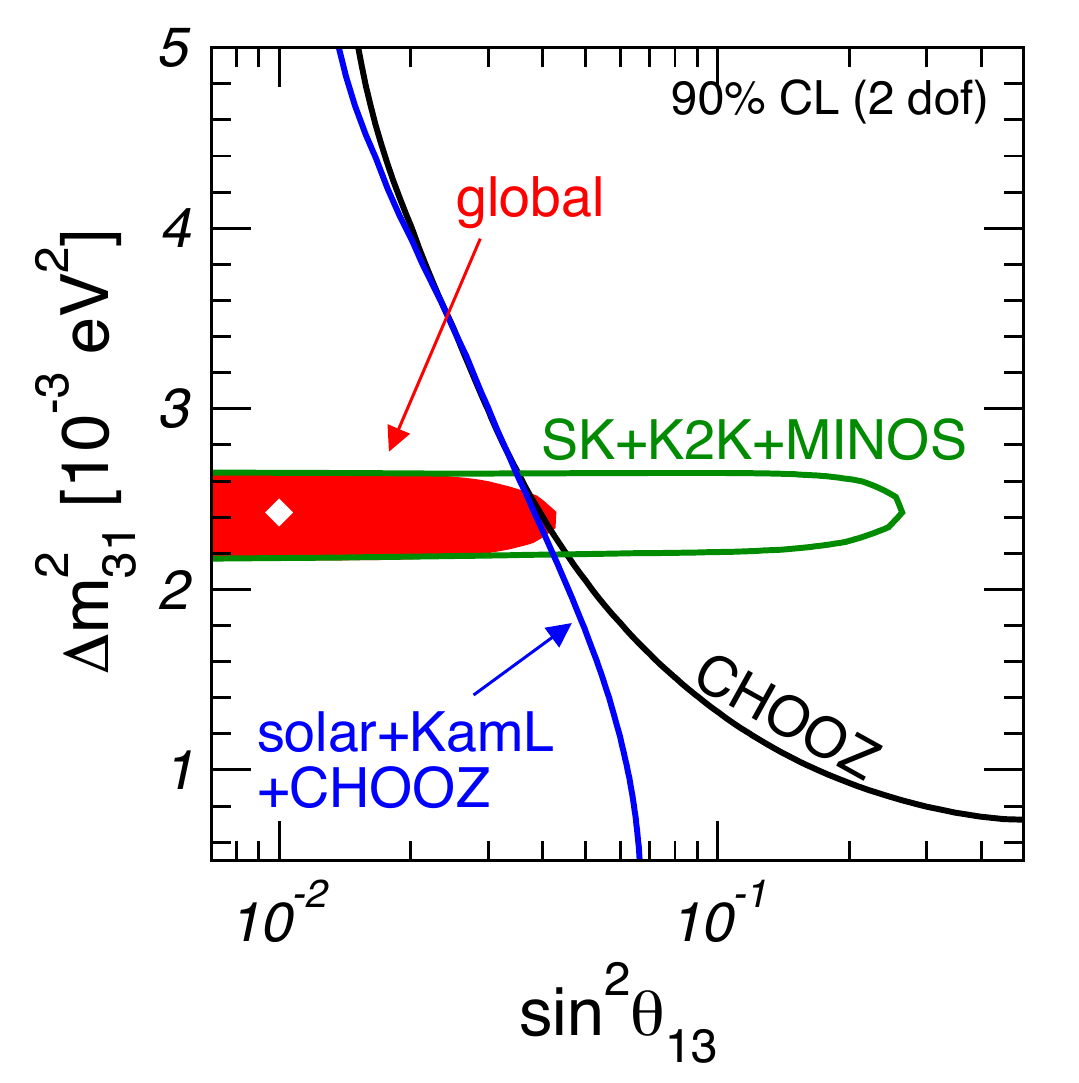} \hspace{0cm}
\raisebox{20pt}[0pt]{\includegraphics[width=0.54\textwidth]{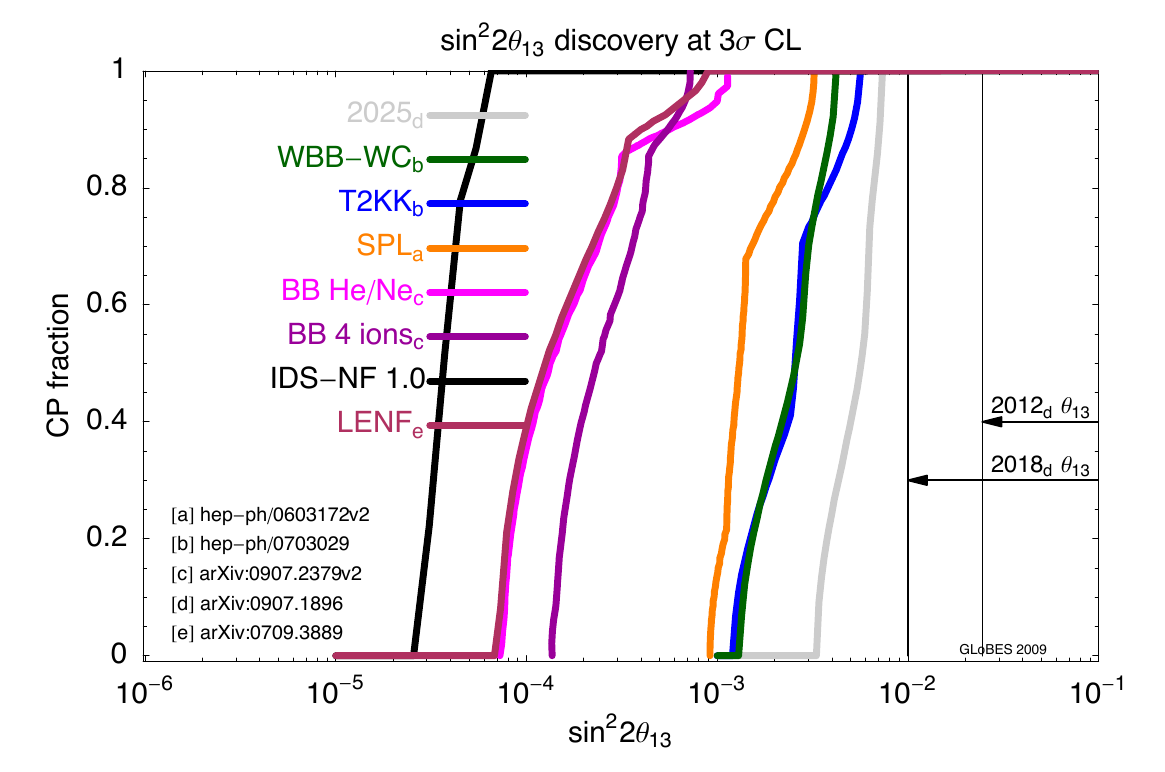}}
\end{center}
\hspace{3.7cm} (a) \hspace{6.9cm} (b) 
\caption{Result of a global fit of $\theta_{13}$ (a). Future prospects on the determination of $\theta_{13}$ (b).}
\label{fig:13}
\end{figure}

The mixing angle $\theta_{13}$ has not been measured yet, but both direct and indirect bounds have been obtained from the CHOOZ and Minos experiments, mentioned above, and from the analysis of subleading effects in the atmospheric and solar neutrino experiments.  The result of a global fit on $\theta_{13}$ are shown in \Fig{13}a~\cite{Schwetz:2008er}. 

The determination of $\theta_{13}$ is important for several reasons. It offers an handle on the origin of the neutrino (and quark) masses and mixing angles. In particular, it allows to discriminate among different flavour models. And it is important for phenomenology, as it is crucial in the study of leptonic CP-violation, supernova signals, and subleading effects, for example in $\nu_\mu\leftrightarrow\nu_\tau$ transitions at the $\dm{23}$ oscillation frequency. From the experimental point of view, a rich experimental program is available. Several terrestrial experiments are running or have been planned using different techniques: conventional beams obtained from pion decays, so called ``beta-beams'', obtained from the beta decay of radio-active ions circulating in a storage ring with
long straight sections, and neutrino factory beams, obtained from the decay of muons also circulating in a storage ring. A summary of the prospects on the $\theta_{13}$ determination are shown in \Fig{13}b for different values of the experimental parameters~\cite{Bernabeu:2010rz}. The figure uses the GLoBES package~\cite{Huber:2004ka,Huber:2007ji}. References for the single experiments are shown in Figure. 

\bigskip

Let us now discuss the determination of the sign of $\dm{23}$. I remind that this parameter determines the pattern of neutrino masses, enters the analysis of supernova neutrino signals and of long baseline terrestrial neutrino experiments, and determines the possibility to measure neutrinoless double beta decay (see below). This parameter can be determined in the presence of matter effects. Let us consider the three neutrino effective Hamiltonian for propagation in matter and let us take the $\dm{12}=0$ limit for simplicity ($\dm{12}$ effects are subleading in the experiments meant to measure $\sign(\dm{23})$): 
\begin{equation}
\label{eq:sign}
H_{\text{eff}} = 
\frac{1}{2E}
\left[U
\begin{pmatrix}
0 & & \\
& 0 & \\
& & \dm{23}
\end{pmatrix}U^\dagger
\pm
\begin{pmatrix}
2EV & & \\
& 0 & \\
& & 0
\end{pmatrix}
\right] .
\end{equation}
The relative sign between the two terms on the RHS depends on whether neutrino or antineutrino oscillations are considered. In the Earth, the MSW potential is positive, $V > 0$. We therefore easily see that if $\dm{23} > 0$ the resonant enhancement of oscillations can take place for neutrinos but not antineutrinos, whereas if $\dm{23} < 0$ the enhancement takes place for antineutrinos. This offers an handle to measure $\sign(\dm{23})$. The neutrino energy should be of the order of the resonant energy, say $10\GeV$, as determined by $|\dm{23}|$ and $V$. Moreover, a long baseline is needed, so that the oscillation phase is not small. I remind in fact that for small oscillation phases $\phi$, the oscillating factor in the probaility can be approximated as $\sin \phi \approx \phi$ and matter effects cancel (see \eq{boh} and below). Finally, the effect shows up in the  $\nu_e\leftrightarrow\nu_{\mu,\tau}$ channel. This can be seen from \eq{sign} by observing that in the limit $\dm{12} = 0$ the $\theta_{12}$ rotation in the PMNS matrix is not physical and $U$ can be approximated with a 23 rotation by the angle $\theta_{23}$. In this limit, matter effects do not affect the $\nu_\mu\leftrightarrow\nu_\tau$ oscillations. The prospects for the measurement of the sign of $\dm{23}$ are summarized in Fig~23 of~\cite{Bernabeu:2010rz}. 

\bigskip

Let us now discuss the possible determination of the CP-violating phase $\delta$. We do not need to stress the importance of investigating whether CP-violation is present not only in the quark sector, but also in the lepton sector. On top of that, leptonic CP-violation could explain the origin of the Baryon asymmetry in the universe through the leptogenesis mechanism. Neutrino oscillation experiments offer the possibility to study leptonic CP-violation associated to the CP phase $\delta$, which is certainly physical, whether neutrinos are Dirac or Majorana. The Majorana phases $\alpha$, $\beta$, if physical, cannot be accessed by oscillation experiments. The CP-violating phase $\delta$ can determine a difference between the neutrino and antineutrino oscillation probabilities: 
\begin{equation}
\label{eq:CP}
\begin{aligned}
P(\nu_{e_i}
\rightarrow
\nu_{e_j})
& = P(\overline{\nu}_{e_j}
\rightarrow
\overline{\nu}_{e_i})
= P_{\text{CPC}}
+ P_{\text{CPV}} \\
P(\overline{\nu}_{e_i}
\rightarrow
\overline{\nu}_{e_j})
& = P(\nu_{e_j}
\rightarrow
\nu_{e_i})
= P_{\text{CPC}}
- P_{\text{CPV}}
\end{aligned}
\end{equation}
(see also \eq{3nuosc}). At accelerators experiments aiming at a measurement of such difference, due to the smallness of $\dm{12}/|\dm{23}|$ and $\theta_{13}$, we can approximate
\begin{align}
\label{eq:CPC}
& 
\left.
\begin{aligned}
P(\nu_\mu\leftrightarrow
\nu_\tau)_{\text{CPC}}
& \approx\sin^2\theta_{23}
\sin^2\frac{\dm{23}L}{4E} \\
P(\nu_e\leftrightarrow
\nu_\mu)_{\text{CPC}}
& \approx\sin^2\theta_{23}
\sin^22\theta_{13}
\sin^2\frac{\dm{23}L}{4E} \\
P(\nu_e\leftrightarrow
\nu_\tau)_{\text{CPC}}
& \approx\cos^2\theta_{23}
\sin^22\theta_{13}
\sin^2\frac{\dm{23}L}{4E}
\end{aligned}
\right\}
+\dm{\text{SUN}}\text{ corr.} 
\\[2mm]
\label{eq:CPV}
& \;\; P_{\text{CPV}} = 
\pm\cos\theta_{13}
\sin2\theta_{12}
\sin2\theta_{23}
\sin2\theta_{13}
\sin\delta\,
\sin\frac{\dm{12}L}{4E}
\sin^2\frac{\dm{23}L}{4E}.
\end{align}
The formulas above show that CP-violation has a chance to show up in $\nu_e\leftrightarrow\nu_\mu$ oscillations~\cite{Dick:1999ed}. First of all, two out of the three angles entering the CP-violating part of the probability in \eq{CPV} are large (unlike the quark mixing angles). If the baseline of the experiment is large enough, the term oscillating with the atmospheric frequency is also of order one. If the phase $\delta$ is not too small, the CP-violating part of the probability is then only suppressed by $\sin2\theta_{13}$ and the solar phase $\sin (\dm{12}L/(4E))$, which are not necessarily too small. On top of that, the CP-conserving part of the $\nu_e\leftrightarrow\nu_\mu$ probability is suppressed by two powers of $\sin2\theta_{13}$, whereas the CP-violating part is suppressed by only one power. This means that the smaller $\sin2\theta_{13}$, the larger is the asymmetry between the probabilities in the neutrino and antineutrino channel,
\begin{equation}
\label{eq:asy}
a_{\text{CP}}
=\frac{P(\nu_e\rightarrow\nu_\mu)
-P(\overline\nu_e
\rightarrow\overline\nu_\mu)}
{P(\nu_e\rightarrow\nu_\mu)
+P(\overline\nu_e
\rightarrow\overline\nu_\mu)}
\propto
\frac{1}{\sin2\theta_{13}+
\text{ corr.}} .
\end{equation}
A smallish $\theta_{13}$ is therefore not necessarily a curse for CP-violation~\cite{Dick:1999ed}. On the one hand, the total number of events decreases with $\sin^22\theta_{13}$, and therefore the statistical error on the measurement of the asymmetry increases as $\delta a \sim 1/\sqrt{N} \propto 1/\sin 2\theta_{13}$, where $N$ is the average number of events. On the other hand, the asymmetry signal also increases with $1/\sin2\theta_{13}$. The statistical significance of the measurement, $\delta a/a$, is therefore approximately constant~\cite{Romanino:1999zq}. Such a behaviour cannot hold for an arbitrarily small value of $\theta_{13}$, of course. This is indeed the case for two reasons: i) the corrections in \eq{asy} become dominant compared to the $\sin2\theta_{13}$ term and ii) the number of expected events may become smaller than one.  

Experimentally, the measurement of CP-violation is complicated by the fake sources of neutrino-antineutrino asymmetry. In particular, one has to consider the CP-asymmetry of the source, which typically does not emit the same number of neutrinos and antineutrinos, the CP-asymmetry of the Earth, made of matter and not antimatter, through which the neutrinos travel, and the CP-asymmetry of the target. A measurement of CP-violation therefore requires a good knowledge of the initial neutrino fluxes, of the Earth (electron) density profile, and of the neutrino cross sections. To cope with such difficulties, it would be useful to have a measurement of the energy spectrum, two baselines, and to measure more than a single oscillation channel. Neutrino factories are especially suited for measuring CP-violation, as the neutrino flux is very high and quite pure. The prospects for the measurement of the phase $\delta$ are summarized in Fig~22 of~\cite{Bernabeu:2010rz}. 

\subsection{Supernova neutrinos}

Supernova neutrinos are emitted during the core collapse of type-II supernovas. Their study can i) provide further informations on the neutrino parameters, modulo the uncertainties on the spectrum and intensity of the source, ii) probe the physics of the collapse, and iii) constrain exotic neutrino transitions, such as oscillations into sterile neutrinos. 

Type-II supernovas originate from the collapse of large stars. The burning process produces heavier and heavier elements in their core. If the star is large enough, the gravitational pressure becomes too large to be stood by the core, and leads to the collapse of the atomic structures. The core, which before collapse has a radius $R\sim 8000\,$km, a density $\rho\sim 10^9\,\text{g/cm}^3$ and a temperature $T \sim 0.7\MeV$, shrinks to a proto-neutron star formed by nuclear matter with $R\sim 30\,$km, $\rho\sim 3\cdot 10^{14}\,\text{g/cm}^3$, $T \sim 30\MeV$. In the process, an impressive amount of energy, $E \sim 3\cdot 10^{53}\,$erg, corresponding essentially to the gravitational energy released, is emitted. Only about 0.01\% of this energy goes into light, about 1\% goes into kinetic energy, and the remaining 99\% is emitted through neutrinos. The neutrino emission is not instantaneous, however. The matter density in the proto-neutron star is so high that the neutrino mean free path is of the order of 10$\,$cm. The time it takes to the neutrinos to diffuse out is then $t_\text{diff} \sim 3R^2/\lambda \sim 10\,$sec. A handful of nupernova neutrinos where detected when the supernova SN 1987A exploded in the Magellanic Cloud, 50$\,$kpc away, in 1987. The time distribution of the neutrino events confirmed the qualitative success of the picture above. 

The observation of the neutrino emission constrains the possibility of invisible, or faster escape channels for the energy to be released. One such example is neutrino oscillations into sterile neutrinos. If the oscillation rate was large enough, sterile neutrinos, which do not interact with matter and would not be trapped inside the core, would immediately escape, carrying away the neutrino energy. A strong bound on a possible active-sterile mixing angle follows, $\sin^2 2\theta_s \lesssim 10^{-8}$, which can be evaded if the sterile neutrino mass is small enough~\cite{Mikheev:1987hv,Shi:1993ee,Dolgov:2000jw,Keranen:2004rg,Nunokawa:1997ct}. Such limits are particularly interesting~\cite{Cacciapaglia:2002qr,Cacciapaglia:2003dx} in the case of neutrinos from extra-dimension~\cite{Dienes:1998sb,ArkaniHamed:1998vp,Dvali:1999cn,Barbieri:2000mg,Mohapatra:2000wn,Lukas:2000wn,Lukas:2000rg,DeGouvea:2001mz}. Other invisible channels constrained by the observation of supernova neutrinos are the conversion into axions or into KK gravitons in large extra dimension scenarios. 

Supernovas in our galaxy are expected to explode with an uncertain, but not very exciting rate of about one every 30 years or more. However, if such an event took place, the present neutrino detectors would gather an impressive amount of data. A supernova 10$\,$kpc away would produce about 8000 neutrino events in SK, 800 in SNO, and 330 in KamLAND, thus allowing a detailed study of the flavour, energy, and time spectrum of the neutrinos reaching us. The distortions of such spectra compared to the expectations in the absence of oscillations (which have a significant degree of uncertainty) might provide information on $\theta_{13}$ and the sign of $\dm{23}$~\cite{Lunardini:2000sw,Barger:2005it}. 

\subsection{Anomalous anomalies}

While the three neutrino oscillation picture consistently and precisely explains an impressive amount of experimental data, the results of the LSND experiment~\cite{Athanassopoulos:1997pv,Aguilar:2001ty} do not fit in the picture. Using a neutrino beam from pion decay detected in a scintillator, such an experiment found an evidence of $\overline{\nu}_\mu \to\overline{\nu}_e$ transitions that, if interpreted in terms of neutrino oscillations, would require a squared mass difference larger than the atmospheric one, $\dm{\text{LSND}} > \dm{\text{ATM}}$. Such a third squared mass difference would require the introduction of a fourth light neutrino. As we have seen in the introduction, the number of light ``active'' neutrinos (i.e.\ with the gauge interactions of standard neutrinos) is bound by the measurement of the $Z$ boson width to be three. The forth neutrino should then be sterile. Such an interpretation poses a number of problems. From the theoretical point of view, in order to account for a light sterile neutrino one should explain while an explicit mass term for it, not forbidden by the electroweak symmetry (unlike the one for active neutrinos), would be absent or extremely small. This can be however accounted for by an appropriate symmetry. Moreover, even if the presence of a fourth neutrino, it is not easy to fit the observed anomaly~\cite{Peres:2000ic,Strumia:2002fw,Maltoni:2002xd} because of the bounds from the Karmen~\cite{Church:2002tc} and Bugey~\cite{Declais:1994su} experiments. The LSND anomaly is being tested by MiniBOONE, which uses about the same value of $L/E$, but with $\ord{10}$ larger values of $L$, $E$. MiniBOONE can run both in a neutrino and antineutrino mode. The present situation is the following. The neutrino run excludes the LSND signal at more than 90\% confidence level (it observes an anomaly, but at the wrong value of $L/E$, and in the low energy region that is more sensitive to the backgrounds)~\cite{AguilarArevalo:2008rc}. The antineutrino run, on the other hand, seems to find an excess compatible with LSND~\cite{AguilarArevalo:2009xn}. 

\subsection{Beyond oscillations}
\label{sec:beyond}

We now discuss the bounds and prospects of determination of the neutrino parameters that cannot be probed with oscillation experiments, $m_\text{lightest}$, and the Majorana phases $\alpha$, $\beta$ (assuming they are physical). 

As said, the determination of the squared mass differences $\dm{23}$ and $\dm{12}$ does not determine the absolute value of neutrino masses. On the other hand, the latter can be obtained from the additional knowledge of $m_\text{lightest}$. We have indeed $m^2_1 = m^2_\text{lightest}$, $m^2_2 = \dm{12} + m^2_\text{lightest}$, $m^2_3 = m^2_\text{lightest} + \dm{12} + \dm{23}$ in the case of normal hierarchy and $m^2_3 = m^2_\text{lightest}$, $m^2_2 = m^2_\text{lightest} - \dm{23}$, $m^2_1 = m^2_\text{lightest} - \dm{12} - \dm{23}$ in the case of inverse hierarchy (in which case $\dm{23} < 0$). In principle, $m_\text{lightest}$ can have any value. If $m_\text{lightest} \ll (\dm{12})^{1/2}\sim 0.01\eV$, the three neutrinos have masses $m_3 \approx |\dm{23}|^{1/2}\sim 0.05\eV$, $m_2 \approx (\dm{12})^{1/2}\sim 0.01\eV$, $m_1= m_\text{lightest} \ll m_2$ in the normal hierarchy case and $m_1 \approx m_2 \approx |\dm{23}|^{1/2}\sim 0.05\eV$, $m_3 = m_\text{lightest} \ll m_{1,2}$ in the inverse hierarchy case (in which case it actually suffices to assume $m_\text{lightest} \ll |\dm{23}|^{1/2}\sim 0.05\eV$). If $m_\text{lightest} \gg |\dm{23}|^{1/2}\sim 0.05\eV$, the three neutrinos are approximately degenerate, $m_1 \approx m_2 \approx m_3 \approx m_\text{lightest}$. 

\bigskip

\begin{figure}
\begin{center}
\includegraphics[width=0.54\textwidth]{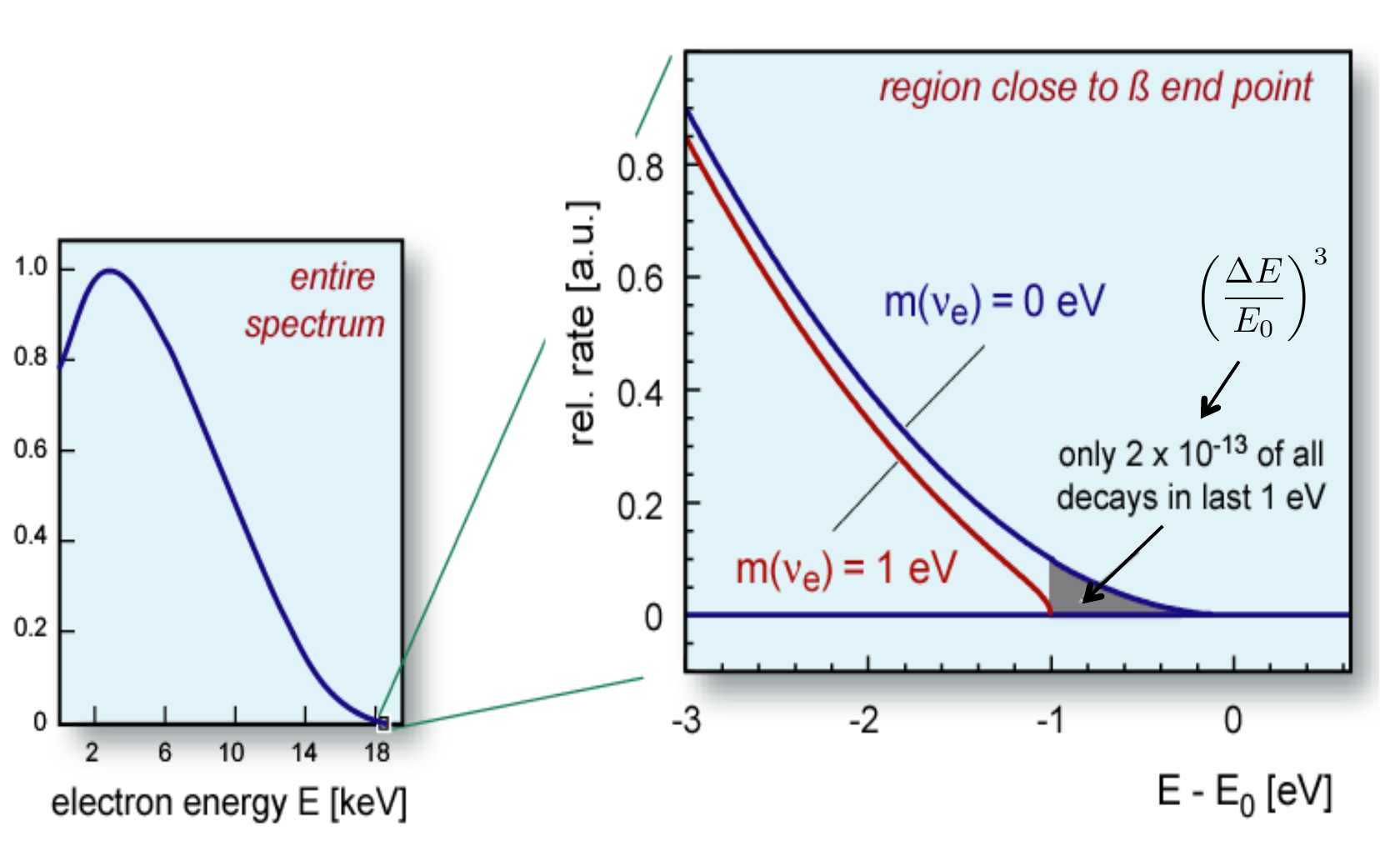} \hspace{0cm}
\raisebox{0pt}[0pt]{\includegraphics[width=0.44\textwidth]{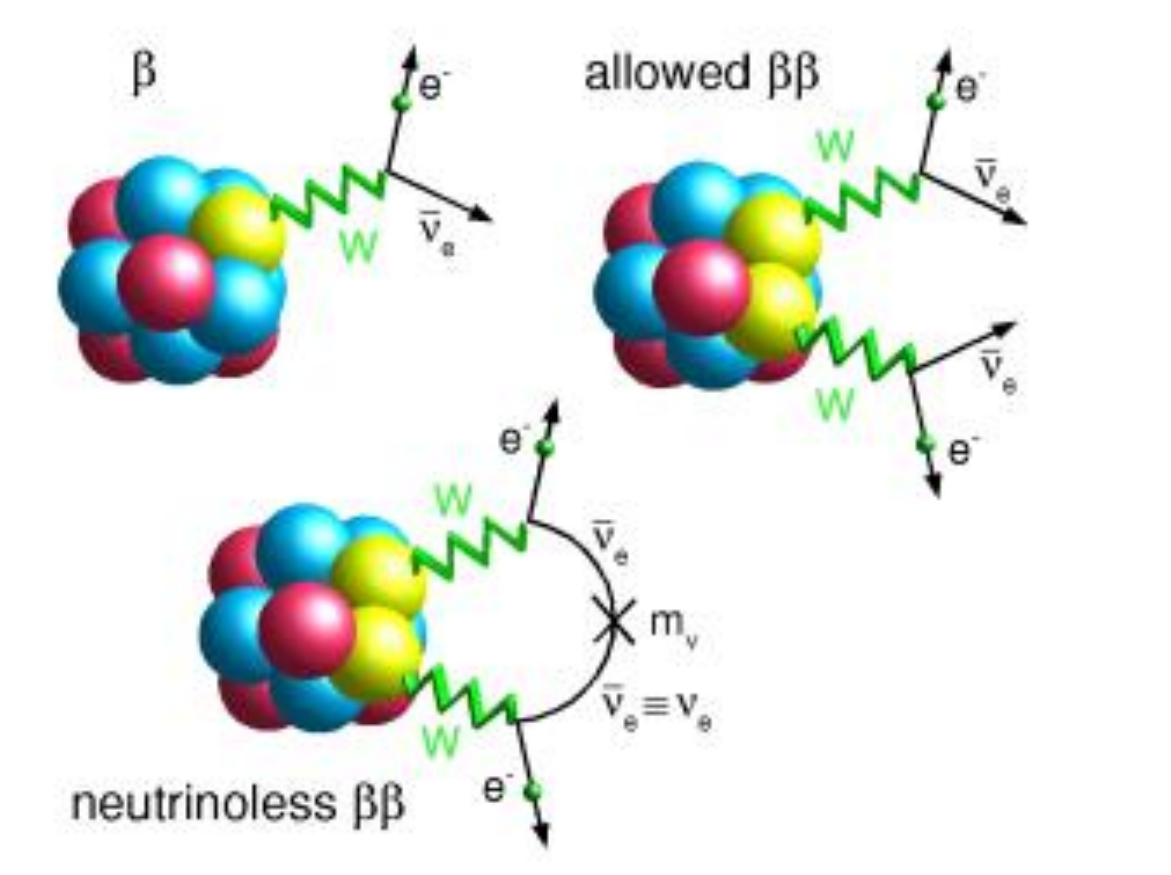}}
\end{center}
\hspace{4cm} (a) \hspace{7cm} (b) 
\caption{Modification of the beta decay spectrum in the presence of a non-vanishing neutrino mass (a). Different beta decay processes and their microscopic mechanisms (b).}
\label{fig:beta}
\end{figure}

Beta decay experiments exploit the fact that a non vanishing neutrino mass modifies the endpoint of the electron spectrum in beta decays $(A,Z) \to (A, Z+1) + e^- + \overline{\nu}_e$, where $A$ and $Z$ are the mass and atomic number of the decaying atom. This is a purely kinematical effect illustrated in \Fig{beta}a (taken from~\cite{Angrik:2005ep}). The tritium decay ${}^3\text{H} \to {}^3\text{He} + e^- + \overline{\nu}_e$ is often used for this purpose. The decay spectrum depends in general on the composition of $\nu_e$ in terms of the three mass eigenstates  and on their masses. Given that the present sensitivities are larger than $|\dm{23}|^{1/2}$, the spectrum only depends on the combination $(m^\dagger m)_{ee}$, where $m$ is the light neutrino mass matrix is the flavour basis:
\begin{gather}
\label{eq:decayspectrum}
\frac{dN}{dE}\propto
\sum|U_{eh}|^2\Gamma(m^2_h,E)
\approx
\Gamma(m^2_{\nu_e},E) , \\[1mm]
\label{eq:mmee}
m^2_{\nu_e} \equiv (m^\dagger m)_{ee}
=|U_{eh}|^2m^2_h
=c^2_{13}
(m^2_1c^2_{12}
+m^2_2s^2_{12})
+m^2_3s^2_{13} .
\end{gather}
The present bound from the Mainz~\cite{Kraus:2004zw} and Troitsk~\cite{Lobashev:2001uu} experiments is  in the degenerate neutrino regime, in which $m_{\nu_e} \approx m_\text{lightest}$, and give $m_\text{lightest} < 2.3\eV$. The Katrin experiment~\cite{Osipowicz:2001sq} promises to lower the sensitivity down to $0.2\eV$. 

\bigskip

Another handle on the absolute value of neutrino masses is provided by neutrinoless double beta ($\nuless$) decay, $(A,Z) \to (A,Z+2) + 2e^-$. As mentioned earlier, $\nuless$ signals lepton number violation and is induced by a Majorana neutrino mass term (see \Fig{beta}b) at a rate $\Gamma \propto |m_{ee}|^2 \langle Q \rangle ^2$, where $\langle Q\rangle$ is the matrix element of the hadronic part of the operator inducing the decay and 
\begin{equation}
\label{eq:mee}
m_{ee}
=U_{eh}^2m_h
=c^2_{13}
(m_1c^2_{12}+m_2s^2_{12}e^{2i\alpha})
+m_3s^2_{13}e^{2i\beta'} 
\end{equation}
is the 11 element of the light neutrino mass matrix in the flavour basis, with $\beta' = \beta-\delta$. The $\nuless$ rate therefore probes both the absolute scale of neutrino masses and the Majorana phases $\alpha$, $\beta$. 

In order to measure the $\nuless$ decay, a nucleus $(A,Z)$ for which the beta decay, but not the double beta one, is kinematically forbidden, is needed. It is then possible to discriminate the neutrinoless decay from the standard two neutrino decay $(A,Z) \to (A,Z+2) + 2e^- + 2\nu_e$. Indeed, the latter has a continuous spectrum for the sum of the energies of the two electrons, with endpoint at the $Q$-value of the decay, while in the former the energy of the electrons must coincide with the $Q$ value. If the energy resolution of the electron energy measurement is precise enough, it is then possible to exclude most of the events due to the standard two neutrino decay. The determination, or bound, one obtains is however plagued by the $\ord{50\%}$ uncertainty associated to the matrix element $\langle Q\rangle$. The Heidelberg-Moscow collaboration, using the $\mbox{}^{76}\text{Ge}\rightarrow \mbox{}^{76}\text{Se}+2e^-$ decay, sets a limit $|m_{ee}|<\ord{1}\times0.4\eV$~\cite{KlapdorKleingrothaus:2000sn}. The claim of a signal has also been reported by a subgroup of the collaboration~\cite{KlapdorKleingrothaus:2004na}. A rich experimental program is available in this field, with prospects of lowering the bound down to a few$\times 10^{-2}\eV$. 

\bigskip

Neutrinos play a role in cosmology through their effect on the Cosmic Microwave Background (CMB) and the formation of Large Scale Structures in the universe (LSS). The effect on CMB is due to the fact that the anisotropies in the photon radiation at decoupling (which takes place at a temperature of about $0.3\eV$) are sensitive to the total radiation density, and in particular to the energy fraction in neutrinos. In turn, the latter is determined by the mere sum of the three neutrino masses, $m_\text{cosmo} = m_1 + m_2 + m_3$, whose knowledge is of course equivalent to the knowledge of $m_\text{lightest}$. The effect on LSS is due to the fact that the free streaming of relativistic non-interacting particles smoothes the density fluctuations leading to the large scale structures observed today. The length scale of the effect depends again on the neutrino masses. 

By fitting the available data on CMB and LSS, it is possible to find an upper bound on $m_\text{lightest}$. However, the latter depends on a number of assumptions (although plausible and consistent) on the cosmological model. It is assumed, for example, that the structures are generated by gaussian adiabatic fluctuations, that the spectral index is constant, that the particle spectrum is the SM one, that the dark matter is cold and the dark energy is accounted for by a non-vanishing cosmological constant. Moreover, we note that the LSS constraint is more powerful but less reliable, as the effect of neutrino masses is larger at smaller scales, where the numerical simulations are more difficult. The bound one obtains at 99\% confidence level is $m_\text{cosmo} < 2.6\eV$ when conservatively using the CMB data only and $m_\text{cosmo} < 0.5\eV$ if the LSS data is also taken into account~\cite{Lesgourgues:2006nd}. 

Besides CMB and LSS, neutrinos also affect Big Bang nucleosynthesis (BBN) and possibly Baryogenesis. The present relative abundance of protons, neutrons, and light elements is determined during BBN by standard inverse beta reactions involving neutrinos at their decoupling temperature $T \sim \MeV$. The Baryon asymmetry in the universe is quantified by the number density of Baryons (minus the negligible density of anti-Baryons), usually normalized to the photon density, $n_B/n_\gamma \approx 6\cdot 10^{-10}$. It is believed that the asymmetry between Baryons and anti-Baryons, $n_B > 0$, originated dynamically during the evolution of the universe. On the other hand, the SM of particle physics cannot account for such a dynamical origin. However, it has been proposed that the Baryon asymmetry could originate from a lepton asymmetry generated by the simplest dynamics underlying the origin of neutrino masses, the see-saw mechanism~\cite{Fukugita:1986hr} (see next Section). More specifically, the idea is that a lepton asymmetry is formed by the CP-asymmetric, out of equilibrium decay of heavy right-handed neutrinos (transformed into a Baryon asymmetry by sphalerons)~\cite{Manton:1983nd,Kuzmin:1985mm,Ambjorn:1990pu}. Although by far not the only one, this is an economical and successful Baryogenesis mechanism that allows, under hypotheses, to relate the single number characterizing the Baryon asymmetry to the neutrino parameters. 

\begin{figure}
\begin{center}
\includegraphics[width=0.3\textwidth]{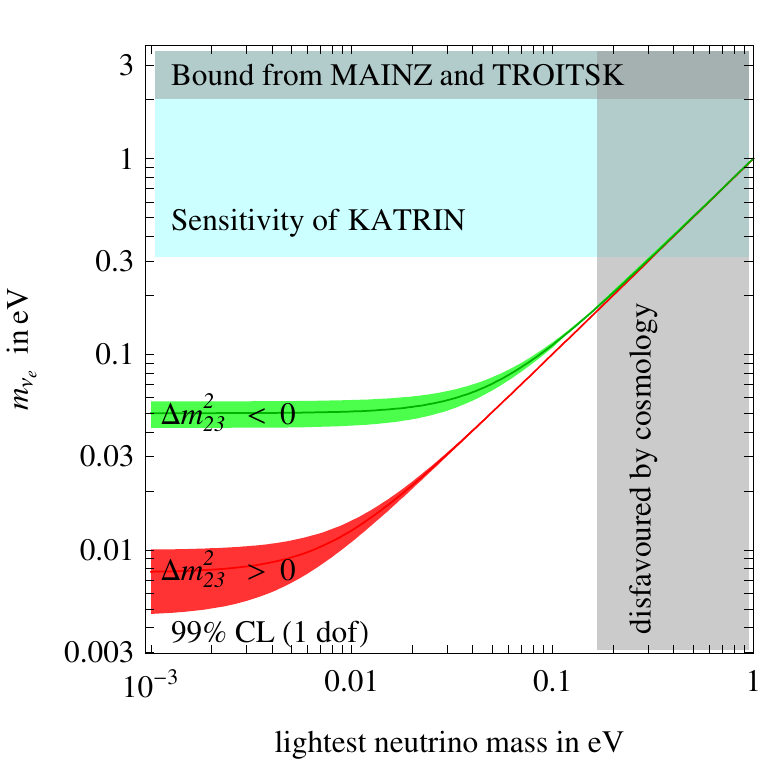} \hspace{0.03\textwidth}
\includegraphics[width=0.3\textwidth]{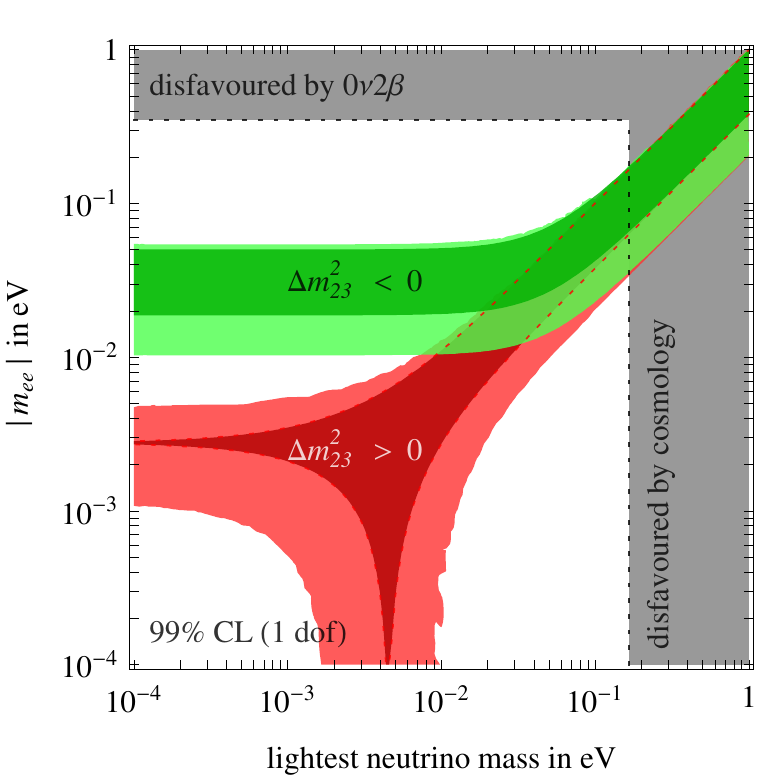} \hspace{0.03\textwidth}
\includegraphics[width=0.3\textwidth]{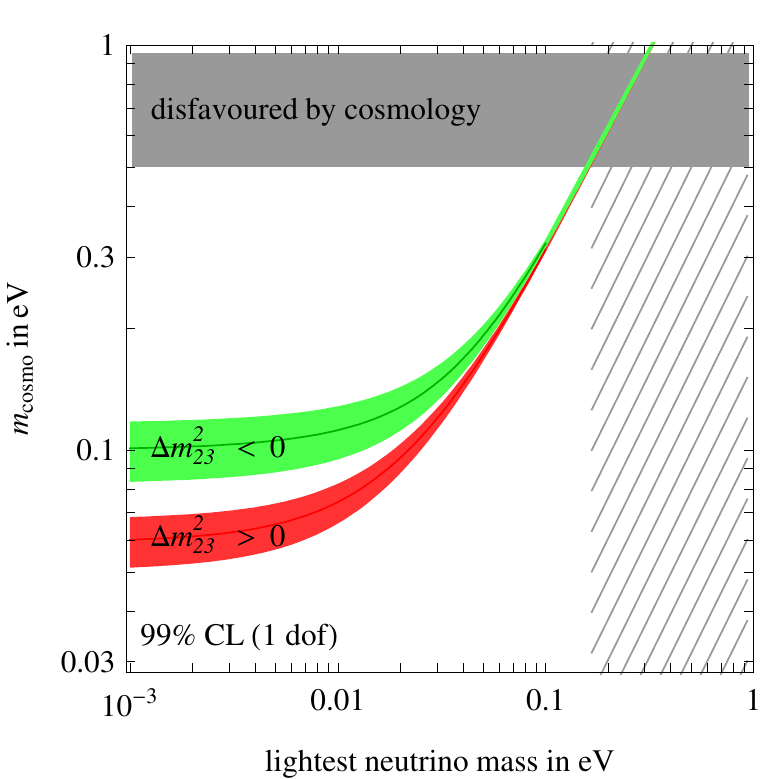} 
\end{center}
\hspace*{2.5cm} (a)
\hspace*{4.6cm} (b)
\hspace*{4.8cm} (c)
\caption{Summary of bounds on $m_\text{lightest}$.}
\label{fig:lightest}
\end{figure}

The summary of theoretical expectations and bounds on $m_\text{lightest}$ is shown in \Fig{lightest}~\cite{Strumia:2006db}. In \Fig{lightest}a the parameter $m_{\nu_e}$ probed by beta decay experiments is plotted as a function of the lightest neutrino mass, taking into account the present uncertainties on $\dm{12}$ and $|\dm{23}|$, for the two signs of $\dm{23}$. The bound from the Mainz and Troitsk experiments are shown together with the expected bound from Katrin. \Fig{lightest}b shows an analogous plot for the parameter probed by $\nuless$ decay. The darker regions correspond to the uncertainty associated to the unknown Majorana phases, with the oscillation parameters fixed at their present central values. The lighter region account for the additional uncertainty on the oscillation parameters. Finally, \Fig{lightest}c shows the situation for the parameter probed by cosmology. 

\section{Theoretical implications}
\label{sec:theo}

After having illustrated the phenomenology associated to neutrino masses and mixings and the determination of the neutrino parameters, we conclude by discussing the theoretical impact of the information that has been gathered so far.

From the theoretical point of view, the relevant information emerging from the data in \eq{parsummary} can be summarized as follows:
\begin{equation}
\label{eq:guidelines}
\begin{gathered}
m_{\nu_i}\ll 174\GeV \\[1mm]
\theta_{23}\sim45^\circ
\text{(= $45^\circ$?)} 
\qquad
\theta_{12}\sim
30^\circ\text{--}35^\circ
\neq45^\circ 
\qquad
\theta_{13}<7^\circ \\[0.5mm]
|\dm{12}/\dm{23}|\approx 
0.035\ll1 .
\end{gathered}
\end{equation}
The most important theoretical guideline is the smallness of neutrino masses. We then have the surprising fact that two out of three mixing angles turn out to be large, unlike what found in the quark sector. In particular, $\theta_{23}$ is compatible with being maximal. While the present uncertainty is too large to draw conclusions, it would be interesting to know whether $\theta_{23}$ is indeed maximal (i.e.\ $45^\circ$ up to small corrections) or just large (i.e. $\ord{1}$). A maximal angle would in fact be an indication of a non-trivial flavour structure~\cite{Altarelli:2010gt}. As for the solar angle, we know that it is definitely not maximal, although compatible with the so called tri-bimaximal prediction~\cite{Harrison:2002er}. The squared mass difference hierarchy implies that the ratio $m_2/m_3$ is about 0.2 or larger, not as small as the typical charged fermion mass hierarchy. In the following, we will concentrate on the first guideline, the smallness of neutrino masses, and its implications for the origin of neutrino masses. 

There is no doubt that neutrino masses are indeed very small compared to the natural scale of fermion masses, the electroweak scale, $v = 174\GeV$: $m_\nu/v \lesssim 10^{-12}$. On the other hand, some of the charged fermion masses  are also quite small compared to $v$, the smallest being the electron mass, suppressed by a factor $m_e/v \approx 0.3 \cdot 10^{-5}$. Still, the smallness of neutrino masses seems to be peculiar. Not only because twelve orders of magnitude are more than five. Also because all the three families of neutrinos are bound to be that light. On the contrary, the suppression of the lightest charged fermion masses seems to be related to the hierarchy among different families, the heaviest families being suppressed compared to the electroweak scale at most by a couple of orders of magnitude. Moreover, there is a compelling explanation for the peculiar smallness of neutrino masses, as we now see. 

\subsection{The origin of neutrino masses}

We have seen in Section~\ref{sec:massterm} that the observed smallness of neutrino masses is not explained at the level of the effective theory below the electroweak scale: the QED and QCD gauge symmetries allow a mass term for both the charged fermions and the neutrinos. Things are different when considering the full SM gauge symmetry $\GSM = \text{SU(3)}_c\times \text{SU(2)}_L\times \text{U(1)}_Y$. It is well known that such a symmetry forbids any fermion mass term, both for charged fermions and neutrinos. In order to see that, it suffices to show that no gauge invariant mass term in the form \eq{massterm} can be written for the left handed fermion fields. The latter transform under the $\text{SU(2)}_L$ gauge symmetry either as doublets, $L_i = (\nu_{iL},e_{iL})^T$, $Q_i = (u_{iL},d_{iL})^T$, or as singlets, $\overline{e_{iR}}$, $\overline{u_{iR}}$, $\overline{d_{iR}}$. Their hypercharges are -1/2, 1/6, 1, -2/3, 1/3 respectively. It is then easy to see that it is not possible to combine any two such left handed fermions in a gauge invariant combination. 

The fact that no fermion mass term is allowed in the limit in which the electroweak symmetry is unbroken can be rephrased by saying that the SM is a ``chiral'' theory. This property might be the very reason why the fermions we observe have a mass so much smaller than, say, the Planck scale: they are protected by the electroweak symmetry. As a consequence, a SM fermion mass has to be proportional to at least one power of $v$. Here is the crucial difference between charged fermions and neutrinos: while the charged fermion mass term arises proportional to one power of $v$, the neutrino mass term needs at least two powers of $v$. Let us see why this is the case and what are the consequences. 

In the SM the electroweak symmetry is broken by the vacuum expectation value (vev) of the Higgs field. The Higgs is a complex scalar field that transforms as a doublet under SU(2)$_L$, $H = (h^+,h^0)$, and has hypercharge 1/2. The Higgs potential is such that the value of the (neutral component of the) field in the ground state does not vanish. Such a vev is denoted by $\langle h^0 \rangle$ and provides the value of the electroweak scale, $\langle h^0 \rangle = v$. The electroweak symmetry is broken because the ground state value of the Higgs is not invariant under electroweak gauge transformations. 

While mass terms for the SM fermions are not allowed, Yukawa interactions with the Higgs are. For example, the electron can interact with the Higgs through the gauge invariant interaction $\lambda_E\overline{e_R} L \, H^\dagger  = \lambda_E \overline{e_R} (e_L h^0 + \nu_L h^+)$, where family indexes have been understood. Once the Higgs field is expressed in terms of the displacement from the ground state value, $h^0 = v + \delta h^0$, a mass term is generated for the electron in the form $m_E \overline{e_R}e_L$, with $m_E = \lambda_E v$. 

The electron Yukawa interaction above has the property of being ``renormalizable''. For our purposes, this means that the coupling in front has a non negative dimension in mass (it is in fact dimensionless). Note that a lagrangian (density) has dimension 4 in energy. Therefore, renormalizable operators have dimension 4 or less. Non-renormalizable terms are instead characterized by coefficients proportional to inverse powers of a mass scale, or cut-off, $\Lambda$. Such terms are thought not to be fundamental but to arise as remnant, effective terms from a more fundamental (possibly renormalizable) theory living at a scale related to $\Lambda$. They are indeed inconsistent (at least perturbatively) at energies larger than $\Lambda$. A dimension $4+n$ non-renormalizable operator is suppressed by $n$ powers of $\Lambda$. 

One can then wonder what is the most general form of the renormalizable Yukawa interactions of the SM fermions with the Higgs, and which is the most general fermion mass term that can be generated at the renormalizable level. The answer is provided by the SM flavour lagrangian
\begin{equation}
\label{eq:flavour}
\begin{alignedat}{5}
\mathcal{L}_\sm^{\text{flavor}} &
&
& = 
\lambda^E_{ij}
\overline{e_{iR}} L_j H^\dagger
&
& + 
\lambda^D_{ij}
\overline{d_{iR}} Q_j H^\dagger
&
& + 
\lambda^U_{ij}
\overline{u_{iR}} Q_j H 
&
& +\text{h.c.} \\
&
&
& = 
m^E_{ij} 
e^c_i e_j
&
& + 
m^D_{ij} 
d^c_i d_j
&
& + 
m^U_{ij} 
u^c_i u_j
&
&+\text{h.c.} 
+ \ldots ,
\end{alignedat}
\end{equation}
with 
\begin{equation}
m^{E}_{ij} = 
\lambda^{E}_{ij} v 
\quad
m^{D}_{ij} = 
\lambda^{D}_{ij} v 
\quad
m^{U}_{ij} = 
\lambda^{U}_{ij} v 
\quad m^\nu_{ij} = 0 .
\end{equation}

We therefore see that, unlike charged fermion masses, neutrino masses do not arise in the SM even after electroweak symmetry breaking, if one sticks to renormalizable interactions. This is a good starting point to understand the smallness of neutrino masses. Of course, we have in the end to account for the fact that neutrino masses are not vanishing. In order to do that, some ingredient has to be added to the SM as a renormalizable theory. While there are certainly several possibilities, we can distinguish two main options. Either the new ingredients live at a scale $\Lambda \gg v$ (the standard example being the addition of heavy right-handed neutrinos giving rise to the see-saw mechanism) or the new ingredients live at a scale $\Lambda\lesssim v$ (the standard example being Dirac neutrinos). Let us consider the two possibilities in turn. 

\subsection{$\Lambda \gg v$}

This case is particularly interesting. It can be described in a model independent way by making use of a central theorem of effective field theory. At the electroweak scale and below, the effect of whatever are the additional heavy degrees of freedom to be added in order to account for neutrino masses, can be described in terms of effective interactions involving only light degrees of freedom and symmetries. In particular, we do not need to know the specific form of the high energy theory in order to parameterize its effect at the electroweak scale in a model independent way. Such effective interactions are non-renormalizable, i.e.\ suppressed by powers of the scale $\Lambda$ at which they arise. We therefore have
\begin{equation}
\label{eq:leff}
\mathcal{L}_{E\ll \Lambda}^\text{eff} =
\mathcal{L}_\text{SM}^\text{ren} +
\mathcal{L}_\text{SM}^\text{NR},
\end{equation}
where $\mathcal{L}_\text{SM}^\text{ren}$ is the renormalizable SM lagrangian and $\mathcal{L}_\text{SM}^\text{NR}$ accounts for the effective interactions. 

The effect at energies $E\ll \Lambda$ of dimension $4+n$ effective interactions arising at the scale $\Lambda$  is suppressed by $(E/\Lambda)^n$, as it can be inferred from simple dimensional analysis. As a consequence, the most relevant effective operators are those with lowest dimension: $4+1$. It turns out that in the SM it is possible to write only one such dimension 5 operator: 
\begin{equation}
\label{eq:dim5}
\mathcal{L}^\text{NR}_\sm =
\frac{h_{ij}}{2\Lambda}
(HL_i)(HL_j) 
+ \text{higher dimension} ,
\end{equation}
where SU(2)$_L$ invariant contractions are understood. Note that separating the coefficient of the operator in \eq{dim5} in a dimensionless numerator $h_{ij}$ and a dimensionful denominator $\Lambda$ is purely conventional. When doing that, we are implicitly identifying with $\Lambda$ the scale at which the degrees of freedom generating the operator live and with $h_{ij}$ the combination of dimensionless couplings, loop factors, etc.\ entering the determination of the operator. 

Once the electroweak symmetry is broken and the neutral Higgs component acquires a vev, a neutrino mass term is generated, 
\begin{equation}
\label{eq:numass}
m^\nu_{ij} = h_{ij} v \times \frac{v}{\Lambda} .
\end{equation}
We therefore see that, unlike charged fermions, neutrino masses turn out in this context to be proportional to two powers of the electroweak symmetry breaking scale, and therefore to be suppressed by a factor $v/\Lambda$ compared to the charged fermion masses, as a consequence of their (unspecified) origin at the high scale $\Lambda$. The smallness of neutrino masses is then understood in terms of the heaviness of the scale $\Lambda$ at which they originate (and at which lepton number if broken). It is also possible to invert the relation in \eq{numass} to obtain an estimate of the scale $\Lambda$ in terms of the measured value of the neutrino masses:
\begin{equation}
\label{eq:Lambda}
\Lambda\sim
0.5\times
10^{15}\GeV h
\fracwithdelims{(}{)}{0.05\eV}{m_\nu} .
\end{equation}
As the coupling $h$  cannot be much larger than 1, $\Lambda$ has to be of the order or smaller than $10^{15}\GeV$. Still, $\Lambda$ might be not too far from the GUT scale. Neutrino masses open in this context an indirect window on scales that could never be probed directly. 

Let us summarize the results of the discussion so far. The smallness of neutrino masses can be  economically understood in a model-independent way in terms of the heaviness of the scale at which lepton number is violated. What makes them special is the fact that they are the only fermions in the SM for which a mass does not arise (after EWSB) from a renormalizable interaction with the Higgs field. They turn out to be Majorana. 

Such an understanding is very appealing, but is based on the fact that, unlike the other charged fermions, neutrinos were not given a right-handed component. In the presence of a right-handed component $\nu_R$, it would be possible to write a renormalizable neutrino Yukawa interaction with the Higgs, $\lambda^N_{ij} \overline{\nu_{iR}} L_j \, H$, providing neutrino masses proportional to the electroweak scale, just as for all the other fermions. The real question might then appear to be: why neutrinos should not have a right-handed components, as all the other fermions? What makes them special from this point of view? 

\begin{figure}
\begin{center}
\includegraphics[width=0.55\textwidth]{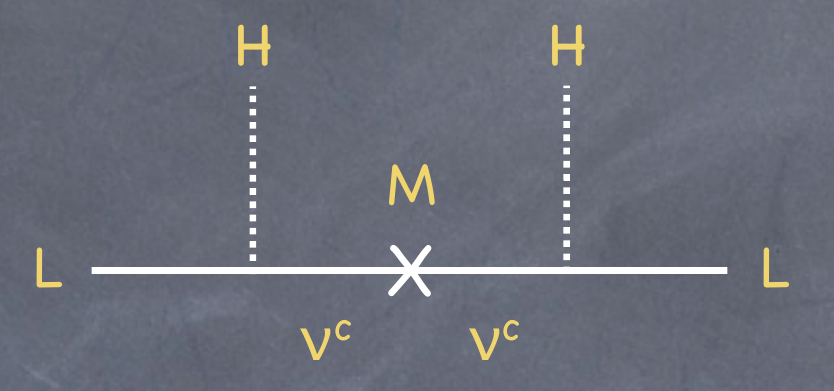} 
\end{center}
\caption{The see-saw diagram.}
\label{fig:seesaw}
\end{figure}

The answer is simple. In order to make the neutrino Yukawa interaction gauge invariant, the right-handed neutrinos should be neutral under all SM interactions\footnote{To be precise, they could also be in a triplet of SU(2)$_L$.}. Then they would be the only fermions for which an explicit mass term in the form in \eq{massterm} would be allowed:
\begin{equation}
\label{eq:nuRmass}
\frac{M_{ij}}{2} \overline{\nu_{iR}}\, \overline{\nu_{jR}} .
\end{equation}
Unlike all other fermions, their mass would not be bound to vanish in the limit in which the electroweak symmetry is unbroken, would not be bound to be proportional to powers of the electroweak scale. Therefore, right-handed neutrinos would be the only fermions for which a mass much larger than $v$ would be allowed. In which case, according to the effective theory theorem mentioned above, their effect at the electroweak scale and below, can be described in terms of the effective interaction in \eq{dim5}. Indeed, neglecting the momentum in the right-handed neutrino propagator in the ``see-saw''~\cite{Minkowski:1977sc,GellMann:1980vs,Yanagida:1979as,Glashow:1979nm,Mohapatra:1979ia} diagram in \Fig{seesaw}a, one obtains the effective interaction in \eq{dim5}, with 
\begin{equation}
\label{eq:seesaw}
\frac{h}{\Lambda}
=
-\lambda_N^T\frac{1}{M}\lambda_N
\quad\Rightarrow\quad
m_\nu = 
-m_{\text{D}}^T
\frac{1}{M}
m_{\text{D}}, \quad \text{where}\quad
m_D = \lambda_N v .
\end{equation}
This is the celebrated see-saw formula. In this context, it turns out to be just an example, probably the simplest, of heavy physics generating the operator in \eq{dim5}. 

It turns out that there are only three possible types of heavy degrees of freedom that can be exchanged at the tree level to generate the operator in \eq{dim5}: the exchange of SM-singlet fermions (type I see-saw, or just see-saw), of a hypercharge -1 SU(2)$_L$ scalar triplet~\cite{Magg:1980ut,Lazarides:1980nt,Mohapatra:1980yp,Schechter:1980gr} (type II see-saw), or of a zero hypercharge SU(2)$_L$ fermion triplet~\cite{Foot:1988aq} (type III see-saw).

\subsection{$\Lambda \lesssim v$}

While the $M \gg v$ option is very appealing, as it provides a solid, economical, compelling understanding of the smallness of neutrino masses, only based on the hypothesis that the new ingredients needed to account for neutrino masses are heavier than the electroweak scale, the possibility that such new ingredients are lighter than $v$ cannot be excluded. In this case, the general effective description we used in the previous case does not hold and each case should be considered separately. 

A paradigmatic example is provided by Dirac neutrinos. As discussed above, such neutrinos have a right-handed component, but their mass terms, which could in principle be as heavy as the Planck scale, is assumed to be much smaller than the neutrino masses themselves. The latter then arise from the Yukawa interaction $\lambda^N_{ij} \overline{\nu_{iR}} L_j \, H$, as for the other fermions, and turn out to be in the Dirac form (as in \eq{nuLnuR} with $m^L = m^R = 0$), with $m_N = \lambda_N v$. 

Such a possibility requires the Majorana mass term for the right-handed neutrinos in \eq{nuRmass} to be smaller than the Planck scale by almost 30 orders of magnitude and the $\lambda_N$ entries to be all smaller than about $10^{-11}$. Lepton number conservation can force $M_{ij} = 0$. We would then only have to cope with very small Yukawas or to account for their smallness with appropriate mechanisms~\cite{Chacko:2003dt,Chacko:2004cz,Davoudiasl:2005ks}. In any case, it is fair to say that additional structure must be added to explain what the simplest option, $\Lambda \gg v$, gives for free.
